\documentclass[fleqn,usenatbib]{mnras}

\usepackage{newtxtext,newtxmath}

\usepackage[T1]{fontenc}

\DeclareRobustCommand{\VAN}[3]{#2}
\let\VANthebibliography\thebibliography
\def\thebibliography{\DeclareRobustCommand{\VAN}[3]{##3}\VANthebibliography}

\usepackage{graphicx}	
\usepackage{amsmath}	
\usepackage{amssymb}	
\usepackage{algpseudocode}
\usepackage{algorithm}
\usepackage{physics}
\usepackage{booktabs}
\usepackage{tikz}
\usetikzlibrary{shapes.geometric, arrows, shapes.misc}

\newcommand{\refsec}[1]{Section~\ref{#1}}
\newcommand{\reffig}[1]{Figure~\ref{#1}}
\newcommand{\reftab}[1]{Table~\ref{#1}}
\tikzstyle{startstop} = [rectangle, rounded corners, minimum width=3cm, minimum height=1cm,text centered, text width=3cm, draw=black, fill=red!30]
\tikzstyle{io} = [trapezium, trapezium left angle=70, trapezium right angle=110, minimum width=1cm, minimum height=1cm, text centered, draw=black, text width=2cm, fill=blue!30]
\tikzstyle{process} = [rectangle, minimum width=3cm, minimum height=1cm, text centered, text width=3cm, draw=black, fill=orange!30]
\tikzstyle{increment} = [rectangle, minimum width=1.5cm, minimum height=1cm, text centered, text width=1.5cm, draw=black, fill=orange!30]
\tikzstyle{decision} = [diamond, minimum width=3cm, minimum height=1cm, text centered, draw=black, fill=green!30]
\tikzstyle{arrow} = [thick,->,>=stealth]
\tikzset{cross/.style={cross out, draw=black, minimum size=2*(#1-\pgflinewidth), inner sep=0pt, outer sep=0pt},
cross/.default={1pt}}

\title[Predicting HR halo clustering using ML]
{A Machine Learning Approach to Correct for Mass Resolution Effects in Simulated Halo Clustering Statistics}

\author[D. Forero-S\'anchez et al.]{
\parbox[t]{\textwidth}{\vspace{-0.8cm}Daniel Forero-S\'anchez,$^{1}$\thanks{E-mail: daniel.forerosanchez@epfl.ch}
Chia-Hsun Chuang,$^{2,3}$\thanks{E-mail: albert.chuang@utah.edu}
Sergio~Rodr\'iguez-Torres$^{4}$,
Gustavo Yepes$^{4}$,
Stefan Gottl\"ober$^{5}$,
Cheng Zhao$^{1}$}
\\
$^{1}$Institute of Physics, Laboratory of Astrophysics,\'Ecole Polytechnique F\'ed\'erale de Lausanne (EPFL), Observatoire de Sauverny, CH-1290 Versoix, Switzerland\\
$^{2}$Department of Physics and Astronomy, University of Utah, Salt Lake City, UT 84112, USA\\
$^{3}$Kavli Institute for Particle Astrophysics and Cosmology, Stanford University, 452 Lomita Mall, Stanford, CA 94305, USA\\
$^{4}$Departamento de F\'isica Te\'{o}rica and CIAFF, M\'{o}dulo 8, Facultad de Ciencias, Universidad Aut\'{o}noma de Madrid, 28049 Madrid, Spain\\
$^{5}$Leibniz-Institut für Astrophysik Potsdam (AIP), An der Sternwarte 16, 14482 Potsdam (Germany)
}

\date{Accepted XXX. Received YYY; in original form ZZZ}

\pubyear{2015}

\begin{document}
\label{firstpage}
\pagerange{\pageref{firstpage}--\pageref{lastpage}}
\maketitle

\begin{abstract}
The increase in the observed volume in cosmological surveys imposes various challenges on simulation preparations. Firstly, the volume of the simulations required increases proportionally to the observations. However, large-volume simulations are quickly becoming computationally intractable. Secondly, on-going and future large-volume survey are targeting smaller objects, e.g. emission line galaxies, compared to the earlier focus, i.e. luminous red galaxies. They require the simulations to have higher mass resolutions. In this work we present a machine learning (ML) approach to calibrate the halo catalogue of a low-resolution (LR) simulation by training with a paired high-resolution (HR) simulation with the same background white noise, thus we can build the training data by matching HR haloes to LR haloes in a one-to-one fashion. After training, the calibrated LR halo catalogue reproduces the mass-clustering relation for mass down to $2.5\times 10^{11}~h^{-1}M_\odot$ within $5~{\rm per~cent}$ at scales $k<1~h\,\rm Mpc^{-1}$. We validate the performance of different statistics including halo mass function, power spectrum, two-point correlation function, and bispectrum in both real and redshift space. Our approach generates high-resolution-like halo catalogues ($>200$ particles per halo) from low-resolution catalogues ($>25$ particles per halo) containing corrected halo masses for each object. This allows to bypass the computational burden of a large-volume real high-resolution simulation without much compromise in the mass resolution of the result. The cost of our ML approach ($\sim 1$ CPU-hour) is negligible compared to the cost of a $N$-body simulation (e.g. millions of CPU-hours), The required computing time is cut a factor of 8.
\end{abstract}




\section{Introduction}
The size of cosmological redshift surveys has been increasing exponentially over the last few decades. The 2dF Galaxy Redshift Survey (2dFGRS) observed $\sim 390~000$ galaxy spectra in an area of $2\times10^3~\rm deg^2$, corresponding to a volume of $0.12~h^{-3}\,\rm Gpc^3$ \citep{Coless2001}. Later, the Sloan Digital Sky Survey\footnote{\url{http://www.sdss.org/sdss-surveys}}'s (SDSS) Baryon Oscillation Spectroscopic Survey (BOSS) observed 1.5 million galaxies within $\sim10^4 \rm deg^2$ \citep{Eisenstein2011, Dawson2013} (a volume of $\sim2.5~h^{-3}\,\rm Gpc^3$). The extended BOSS (eBOSS) expanded these observations through the inclusion of different matter tracers and expanding surveyed redshift range \citep{Dawson2016} resulting in an extended volume of $\sim1.5~h^{-3}\,\rm Gpc^3$. Currently the Dark Energy Spectroscopic Instrument\footnote{\url{http://desi.lbl.gov/}} (DESI) is observing a $14\times10^3~\rm deg^2$ field and is expected to acquire more than 30 million redshifts \citep{Levi2019} in a volume of $\sim20~h^{-3}\,\rm Gpc^3$. When finished, DESI will provide the largest observed cosmological volume available. Meanwhile, future-generation large surveys are already being developed, e.g., 
4MOST\footnote{\url{http://www.4most.eu/}} (4-metre Multi-Object Spectroscopic Telescope, \citealt{deJong2019}),
HETDEX\footnote{\url{http://hetdex.org}} (Hobby-Eberly Telescope Dark Energy Experiment, \citealt{Hill:2008mv}),
PFS\footnote{\url{https://pfs.ipmu.jp}}(Subaru Prime Focus Spectrograph, \citealt{2014PASJ...66R...1T}),
Euclid\footnote{\url{http://www.euclid-ec.org}} \citep{Laureijs:2011gra}, 
and the Roman Space Telescope\footnote{\url{https://roman.gsfc.nasa.gov}} \citep{Spergel:2013tha}.  These surveys aim to investigate the properties of dark energy (DE) through a precise measurement of the Baryon Acoustic Oscillations (BAO) scale and the growth rate of the Universe through Redshift Space Distortions (RSD). In summary, the volume of observed Universe can be expected to grow significantly over the next decades. 

To constrain cosmological parameters using the observed galaxy distribution, we need to compare the distribution with the prediction based on theoretical models. At linear and quasi-linear scales, one can build the models based on perturbation theories. However, to extract cosmological information from small-scale clustering, we need to take into account the nonlinear effects while building the models. This can be achieved by modelling non-linearities with semi-analytical or empirical models that rely on approximations of the structure formation process \citep{Wang2017}. To validate these approximations, make use of numerical solutions, i.e., $N$-body simulations which encode a complete description of gravitational evolution \citep{Alam2021}. However, to provide an accurate and precise prediction, a simulation should have a high resolution to resolve proper haloes hosting the observed galaxies and a large volume to reduce cosmic variance. While large volume simulations may be obtained in a reasonable time by reducing the mass resolution, simulations complying with both these requirements are very computationally expensive.

Due to their high computational costs, $N$-body cosmological simulations have challenged generations of hardware architectures. Despite developments in hardware and software \citep{Springel2005, Ishiyama2009, Habib2016, Potter2017, Garrison2019}, the requirements of modern cosmological surveys -- both in terms of number of realizations and their resolution -- may soon render exact $N$-body computations intractable. Current software and hardware takes upwards of two million CPU hours for a single $1~h^{-1}\rm Gpc$ High Resolution (HR) simulation used in this study, while we require the simulation volume to be much larger than observed volume (e.g. $\sim20~h^{-3}\rm{Gpc}^3$ for DESI) to minimize the theoretical uncertainties.

The increasing availability of large data sets has facilitated the development of Machine Learning (ML) methods, as well as their application to various aspects of science and technology. In particular, ML in the physical sciences has helped overcome great barriers in computation related to forward modelling and simulations in various different fields. Various types of ML techniques, ranging from classical methods such as Kernel Ridge Regression \citep{Bartok2013} to modern deep architectures such as Generative Adversarial Networks (GAN) \citep{Mustafa2019} have been found valuable for their specific research fields. A complete review of applications of ML in physics (and vice-versa) can be found in \citet{Carleo2019}.

Moreover, astronomical and astrophysical research has also seen a significant increase in the applications of ML techniques. For example, Convolutional Neural Networks have been widely used to analyse weak lensing maps \citep{Gupta2018, Ribli2019, Lu2021}. Additionally, other architectures such as Variational Autoencoders have been used in Cosmic Microwave Background data analysis \citep{Yi2020}. Classical ML methods, such as random forests (RF) are also widely used due to their interpretability and capacity, for example in photometric redshift estimation \citep{Carrasco2013, Mountrichas2017}. Within the analysis of the Large Scale Structure of the Universe, ML techniques have been developed to perform BAO reconstruction \citep{Mao2021}, and specially to generate and emulate large cosmological simulations \citep{He2019}. Examples of the latter take advantage of GANs to generate new simulations from a random state \citep{Feder2020} and even increase the resolution of density maps \citep{Li2021, Ni2021}. 

Many of these methods use well-studied neural network architectures that operate on pixelated data. This fact poses challenges in terms of storage size and computational cost of training, due to the necessity of extending these models for three-dimensional inputs and outputs. Therefore, the cosmological volume and the resolution that the network can be trained on is largely limited and high-performance computational facilities and large computing time allocations may still be necessary to train and infer from these models. In this work we present an alternative that does not rely on neural networks on pixelated data, but on individual halo properties and engineered environmental features. This approach allows grid-independent training and inference.

In the next section (\refsec{sec:data}) we introduce the $N$-body simulations our model is based on and describe the halo catalogues extracted from them. In \refsec{sec:build-train} we explain how the training data set is constructed from the base simulations and halo catalogues. We also introduce the feature-set of the different haloes. Next, \refsec{sec:models} describes the ML models used in this work. \refsec{sec:evaluation-metrics} describes the different metrics used to evaluate the performance of our model. \refsec{sec:results} presents the results. 
Finally in \refsec{sec:conclusions} we conclude and state future work directions.

\section{Data}
\label{sec:data}
In this section we introduce the data used for this project and some of the processing steps from the raw simulations to dark matter haloes.
\subsection{UNIT Simulations}
The Universe N-body simulations for the Investigation of Theoretical models from galaxy surveys\footnote{\tt http://www.unitsims.org} \citep[UNIT;][]{Chuang2019} is a set of cosmological $N$-body simulations that aims to provide low-variance clustering statistics from a gravity solver while avoiding the computational cost of a large ensemble of independent simulations. The simulations are based on the method proposed by \citet{Angulo2016}, which shows that a single pair of fixed-amplitude and phase-paired simulations would yield clustering measurements similar to those of 50 averaged independent simulations. More precisely, the pair of simulations has the initial mode amplitudes fixed to the ensemble-averaged power spectrum and the phases of the random field are chosen to be perfectly off-phase such that the noise is cancelled out. Thus, only a small number of simulations is required.

In the present work we use two sets of UNIT simulations evolved from the same initial conditions but with different mass resolutions. High resolution UNIT simulations are obtained using the \textsc{Gadget}  \citep{Springel2005} code with $4096^3$ particles of mass $1.2\times10^{9}~h^{-1}M_{\odot}$ in a volume of $1~(h^{-1}\rm Gpc)^3$. This mass resolution is expected to be consistent with the requirements of current-generation surveys. In particular, the high resolution simulations should resolve haloes hosting Emission Line Galaxies (ELGs), that is, of mass $\sim10^{11}~h^{-1}M_{\odot}$ \citep{Gonzalez2018}. 
Low resolution (LR) simulations are also done with \textsc{Gadget}, but with $2048^3$ particles of mass $9.9\times10^{9}~h^{-1}M_{\odot}$ in the same volume.

From these simulations, HR haloes are extracted using the Friend-of-Friends (FOF) algorithm described in appendix B of \cite{Riebe2013}. The chosen linking length corresponds to 0.2 times the mean inter-particle separation. To build the training set we choose a minimum of 100 HR particles per halo and a minimum of 8 LR particles per halo. The extracted halo catalogues contain a number of halo properties that are used in our experiments as features of the model. These are enumerated in \reftab{tab:halo-properties} along with their definitions.

The cosmological parameters used in the UNIT simulations and through this work are $\Omega_m = 0.3089$, $h = 0.6774$, $n_s = 0.9667$ and $\sigma_8 = 0.8147$  \citep{Planck2016}.

\begin{table*}
\centering
\caption{Table summarising FOF halo properties and their definitions. See appendix B of \citet{Riebe2013} for more details. $G$ is the gravitational constant.}
\begin{tabular}{l|l|l}
\hline
\multicolumn{1}{c|}{\textbf{Name}} & \multicolumn{1}{c|}{\textbf{Halo property}}           & \multicolumn{1}{c}{\textbf{Definition}}                                                                                                           \\ \hline
\texttt{log\_halo\_mass}           & (Log) Halo mass, $\log(M_{\rm halo})$                 & \hspace{3mm}Total mass of the FOF halo. Sum of particle masses within the FOF group.                                                                           \\ \hline
\texttt{delta}                     & Mean overdensity, $\delta$                            & \begin{tabular}[c]{@{}l@{}}Overdensity computed using the density defined by the mass enclosed in a sphere\\ of radius $r_{\rm sph}$.\end{tabular} \\ \hline
\texttt{log\_sigma}                &  (Log) Particle position dispersion within halo, $\log\sigma$ &  \hspace{3mm}Dispersion of simulation particle positions within the FOF group.   \\      \hline
\texttt{log\_sigma\_v}             & (Log) Velocity dispersion within halo, $\log\sigma_v$ & \hspace{3mm}Dispersion of simulation particle velocities within the FOF group. \\\hline
\texttt{log\_r\_sph}               & (Log) Halo radius, $\log r_{\rm sph}$                 & \begin{tabular}[c]{@{}l@{}}Halo size defined as the radius of the sphere containing the same volume as the\\ FOF group.\end{tabular}              \\ \hline
\texttt{log\_spin\_p}              & (Log) Spin parameter $\log\lambda$                    & \hspace{3mm}$\lambda = \frac{JE_{\rm kin}^{1/2}}{GM_{\rm halo}^{5/2}}$ defined by the angular momentum $J$ and the kinetic energy $E_{\rm kin}$.                \\ \hline
\texttt{v}                         & Center of mass speed, $v$.                             & \hspace{3mm}Norm of the center of mass halo velocity $v = \abs{\vb*{v}}$                                                                                      \\ \hline
\end{tabular}
\label{tab:halo-properties}
\end{table*}




Given that we have only a single pair of UNIT simulations, we divide it in training, validation and test sets of sizes of $(0.5\times1\times1)$, $(0.5\times0.5\times1)$ and $(0.5\times0.5\times1)~h^{-3}\rm Gpc^3$ respectively. The training set is used to fit the model parameters, while the validation test is used to optimize the different hyperparameters. Finally the test set is used to evaluate the performance of the model once the hyperparameters are all fixed.

\section{Construction of the training set}
\label{sec:build-train}
To build the training data, we need to match haloes from the low resolution simulation to haloes from the high resolution simulation. In this section we first introduce the feature selection and engineering. Then, we describe how the targets of the model are obtained via one-to-one halo matching.

\subsection{One-to-one halo matching}
\label{sec:halo-matching}
In a general sense, the halo matching procedure assigns a target mass -- taken from a HR halo -- to a LR halo. This task is not trivial, due to the possibility that the LR simulation does not resolve all individual HR haloes. Nonetheless, we aim to find the best one-to-one matching between catalogues of both resolutions. The assignment is designed to take into account two factors. 
\begin{enumerate}
    \item The pair of matched haloes must be spatially close.
    \item The masses of the haloes must be close since the LR haloes considered should be reasonably resolved and we aim at calibrating their masses.
\end{enumerate}

In order to fulfil these conditions we start by identifying the $k_{\rm nb}$ nearest LR neighbours of each HR halo within a distance $r_{\rm match}$ as match candidates using \textsc{Scipy}'s kd-tree implementation \citep{Virtanen2020}. To choose the most appropriate candidate among the pool of LR neighbours we aim to select the one that is closest in terms of both position and mass. While the spatial distance is trivially chosen to be the Euclidean distance, we define the `mass distance' as the relative mass difference:
\begin{equation}
    \Delta_M = \frac{\abs{M_{\rm LR} - M_{\rm HR}}}{M_{\rm HR}},
\end{equation}
where $M_{\rm LR}$ is the mass of the LR candidate and $M_{\rm HR}$ is the mass of the HR halo. We condition the matching on the mass distance fulfilling $\Delta_M < \Delta_M^{\rm th}$. Here we introduce an extra hyperparameter $\Delta_M^{\rm th}$ that allows us to control the completeness of which the HR haloes are matched. We find that larger values of $\Delta_M^{\rm th}$ allow for more HR haloes to be matched but bias the clustering measurements of the corrected catalogue since there are more pairs of HR and LR haloes with very different masses.

\begin{figure}
    \centering
    \begin{tikzpicture}
         
        \node[rectangle,
            draw = gray,
            fill = gray!10!white,
            minimum width = 8cm, 
            minimum height = 1.8cm] (lr) at (0,1) {};
            
        \node[rectangle,
            draw = gray,
            fill = gray!10!white,
            minimum width = 8cm, 
            minimum height = 1.8cm] (hr) at (0,3.5) {};

        \node[rectangle,
            minimum width = 0.2cm, 
            minimum height = 0.2cm] (hrl) at (-3.5,3.9) {HR};

        \node[rectangle,
            minimum width = 0.2cm, 
            minimum height = 0.2cm] (lrl) at (-3.5,1.4) {LR};

        \node[circle,
            fill=gray!60!white,
            minimum width = 0.8cm, 
            minimum height = 0.8cm] (lc1) at (-2.8,1) {};

        \node[circle,
            fill=gray!60!white,
            minimum width = 0.8cm, 
            minimum height = 0.8cm] (lc2) at (-1.8,1) {};

        \node[circle,
            fill=gray!60!white,
            minimum width = 0.8cm, 
            minimum height = 0.8cm] (lc3) at (-0.8,1) {};

        \node[circle,
            fill=gray!60!white,
            minimum width = 0.8cm, 
            minimum height = 0.8cm] (lc4) at (2.8,1) {};

        \node[circle,
            fill=gray!80!white,
            minimum width = 0.8cm, 
            minimum height = 0.8cm] (hc1) at (-2.8,3.5) {};

        \node[circle,
            fill=gray!80!white,
            minimum width = 0.8cm, 
            minimum height = 0.8cm] (hc3) at (2.8,3.5) {};

        \node[circle,
            fill=gray!80!white,
            minimum width = 0.8cm, 
            minimum height = 0.8cm] (hc2) at (1.4,3.5) {};

        \draw [arrow] (lc1) -- (hc1);
        \draw [arrow,
                color=red] (lc4) -- (hc3);
        \draw [arrow,
                color=red] (lc4) -- (hc2);

    \end{tikzpicture}
    
    \begin{tikzpicture}
         
        \node[rectangle,
            draw = gray,
            fill = gray!10!white,
            minimum width = 8cm, 
            minimum height = 1.8cm] (lr) at (0,0) {};
            
        \node[rectangle,
            draw = gray,
            fill = gray!10!white,
            minimum width = 8cm, 
            minimum height = 1.8cm] (hr) at (0,2.5) {};

        \node[rectangle,
            minimum width = 0.2cm, 
            minimum height = 0.2cm] (hrl) at (-3.5,2.9) {HR};

        \node[rectangle,
            minimum width = 0.2cm, 
            minimum height = 0.2cm] (lrl) at (-3.5,0.4) {LR};

        \node[circle,
            fill=gray!60!white,
            minimum width = 0.8cm, 
            minimum height = 0.8cm] (lc1) at (-2.8,0) {};

        \node[circle,
            fill=gray!60!white,
            minimum width = 0.8cm, 
            minimum height = 0.8cm] (lc2) at (-1.8,0) {};

        \node[circle,
            fill=gray!60!white,
            minimum width = 0.8cm, 
            minimum height = 0.8cm] (lc3) at (-0.8,0) {};

        \node[circle,
            fill=gray!60!white,
            minimum width = 0.8cm, 
            minimum height = 0.8cm] (lc4) at (2.8,0) {};

        \node[circle,
            fill=gray!80!white,
            minimum width = 0.8cm, 
            minimum height = 0.8cm] (hc1) at (-2.8,2.5) {};

        \node[circle,
            fill=gray!80!white,
            minimum width = 0.8cm, 
            minimum height = 0.8cm] (hc3) at (2.8,2.5) {};

        \node[circle,
            fill=gray!80!white,
            minimum width = 0.8cm, 
            minimum height = 0.8cm] (hc2) at (1.4,2.5) {};

        \draw [arrow] (lc1) -- (hc1);
        \draw [arrow,
                color=black] (lc4) -- (hc3);
        \draw [arrow,
                color=black] (lc3) -- (hc2);
        \draw (-1.8,0) node[cross,rotate=0,
                            minimum width = 0.6cm,
                            minimum height = 0.6cm,
                            color=red] {};
        \path (current bounding box.north) ++ (0,0.8cm);
    \end{tikzpicture}
    \caption{Halo matching process: under the condition $\Delta_M < \Delta_M^{\rm th}$, haloes in the high-resolution (HR) box find nearest haloes in the low-resolution (LR) box as their matches. \textit{Upper panel:} Arrows indicate the nearest haloes found. Red arrows indicate competitions (multiple HR haloes found the same LR halo as their nearest haloes). \textit{Lower panel:} When a competition occurs, the closer HR halo gets to keep the LR halo and the other HR halo needs to find the next nearest LR halo. At the end, we discard the rest of the LR haloes indicated with the red cross.}
    \label{fig:matching_process}
\end{figure}
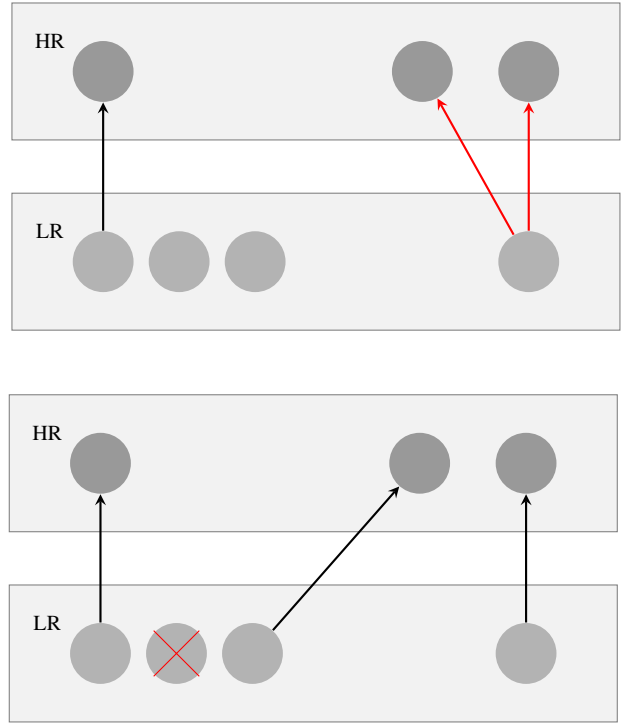

\reffig{fig:matching_process} shows the procedure of halo matching process. During the halo matching, we iterate over all the HR haloes (i.e. haloes with $>100$ particles) and look for their counterparts in the LR halo catalogue. It is possible, however, that multiple HR haloes are assigned to the same LR halo. If a LR halo $L$ has initially been assigned to a HR halo $H_0$, but it is later found that a different HR halo $H_1$ is a better match (i.e. is spatially closer), then $L$ is matched with $H_1$ and $H_0$ is marked as unmatched. We loop multiple times over the whole the HR catalogue to allow the unmatched haloes to be paired with their next-best candidate. In practice, we find 5 iterations to be enough for pairing most of the HR haloes and iterations yield no new pairings.
   
 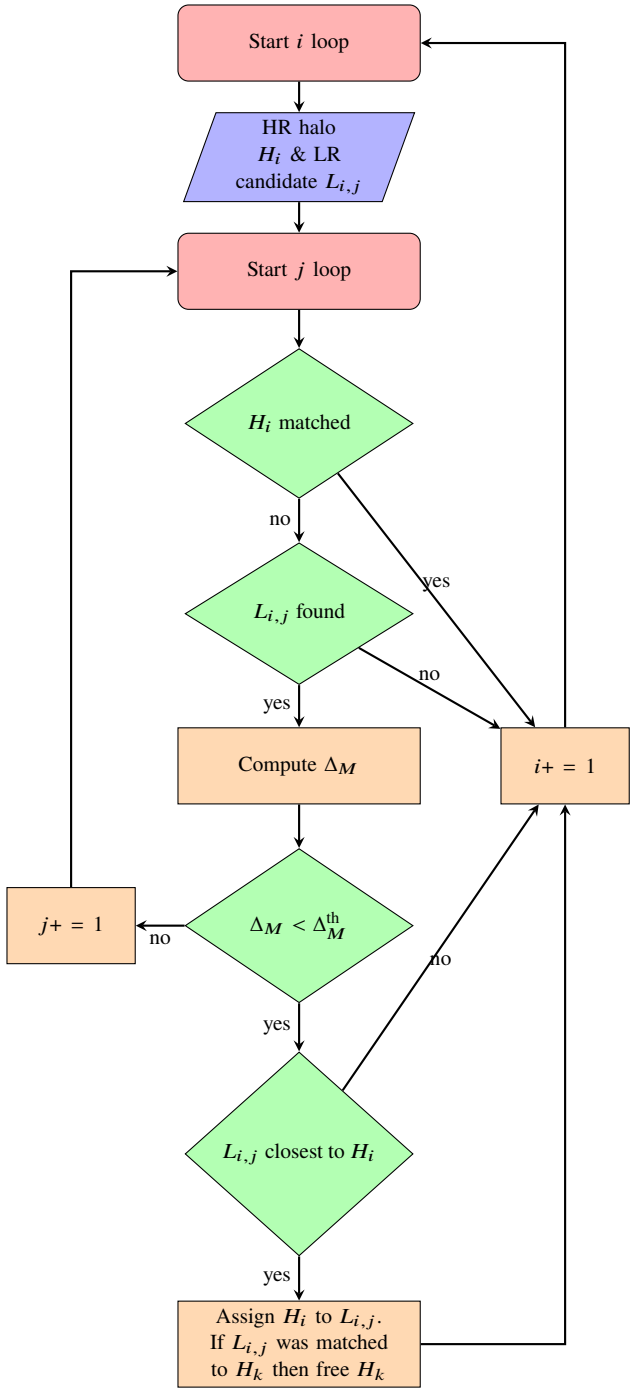
\begin{figure}
    \centering
    \begin{tikzpicture}[node distance=1.5cm]
    \node (starti) [startstop] {Start $i$ loop};
    \node (in1) [io, below of=starti] {HR halo $H_i$ \& LR candidate $L_{i,j}$};
    \draw [arrow] (starti) -- (in1);
    \node (startj) [startstop, below of=in1] {Start $j$ loop};
    \draw [arrow] (in1) -- (startj);
    \node (dec1) [decision, below of=startj, yshift=-0.5cm] {$H_i$ matched};
    \draw [arrow] (startj) -- (dec1);
    \node (dec2) [decision, below of=dec1, yshift=-1cm] {$L_{i,j}$ found};
    \node(proc2) [process, below of=dec2, yshift=-0.5cm] {Compute $\Delta_M$};
    \node (dec3) [decision, below of=proc2, yshift=-0.6cm] {$\Delta_M < \Delta_M^{\rm th}$};
    \node(proc1) [increment, right of=proc2, xshift=2cm] {$i+=1$};
    \draw [arrow] (dec1) -- node[anchor=south] {yes} (proc1);
    \draw [arrow] (dec1) -- node[anchor=east] {no} (dec2);
    \draw [arrow] (proc1) |- (starti);
    \draw [arrow] (dec2) -- node[anchor=east] {yes} (proc2);
    \draw [arrow] (dec2) -- node[anchor=south] {no} (proc1);
    \node(proc3) [increment, left of=dec3, xshift=-1.5cm] {$j+=1$};
    \draw [arrow] (proc2) -- (dec3);
    \draw [arrow] (proc3) |- (startj);
    \node (dec4) [decision, below of=dec3, yshift=-1.5cm] {$L_{i,j}$ closest to $H_i$};
    \draw [arrow] (dec3) -- node[anchor=east] {yes} (dec4);
    \draw [arrow] (dec3) -- node[anchor=north] {no} (proc3);
    \node(proc4) [process, below of=dec4, yshift=-1cm] {Assign $H_i$ to $L_{i,j}$. If $L_{i,j}$ was matched to $H_k$ then free $H_k$};
    \draw [arrow] (dec4) -- node[anchor=east] {yes} (proc4);
    \draw [arrow] (dec4) -- node[anchor=north] {no} (proc1);
    \draw [arrow] (proc4) -| (proc1);
    \end{tikzpicture}    
    \caption{Loop for halo matching. The distance between all $L_{i,j}$ and its match is initialized to infinity. It is possible that there are less than $k_{\rm nb}$ neighbours found within $r_{\rm match}$, in which case the candidate $L_{i,j}$ may not exist. Nearest neighbours are sorted by distance such that $L_{i,j}$ is closer than $L_{i,j+1}$ for all $i,\,j$.}
    \label{fig:halo-match-flow}
\end{figure}

\begin{figure}
    \centering
    \includegraphics[width=\linewidth]{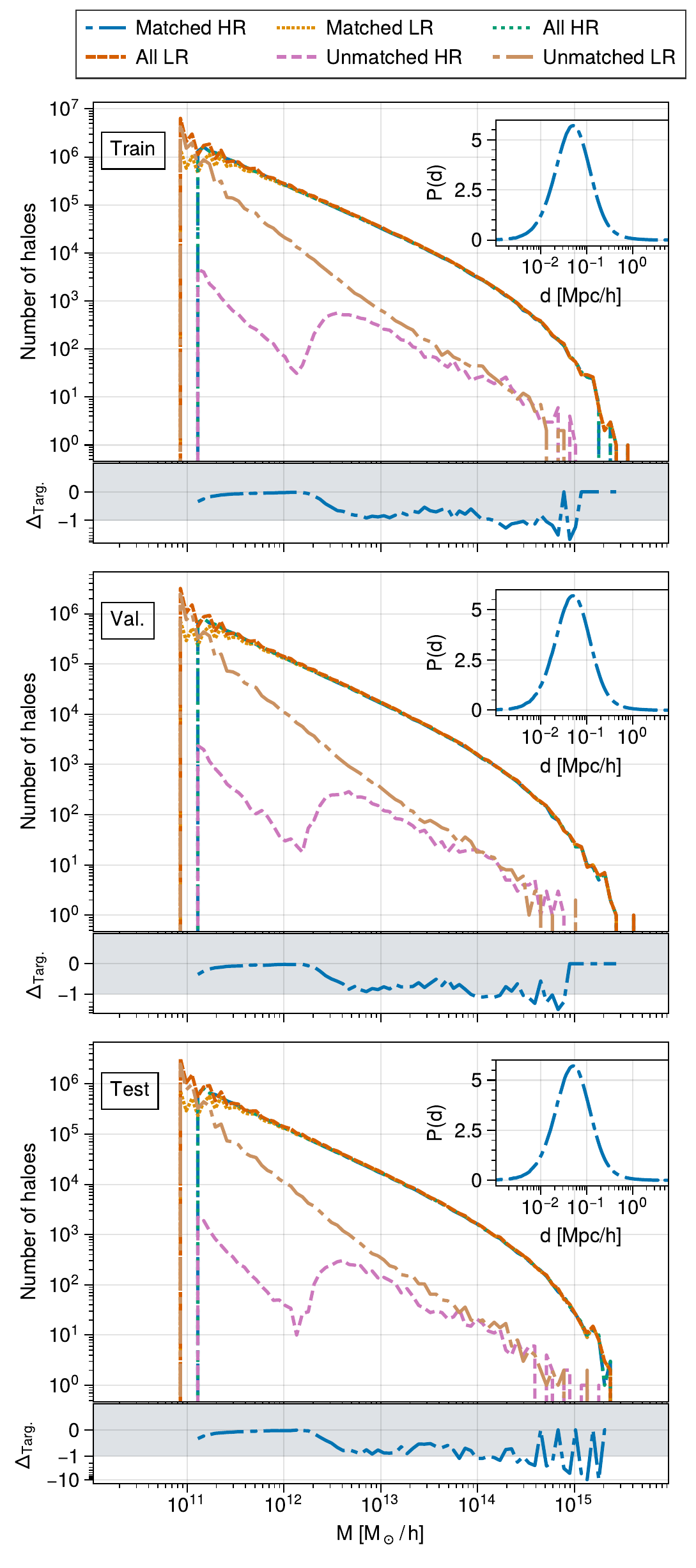}
    \caption{Halo mass functions resulting from the halo matching procedure for the training, validation and test sets. The sub-panels show $\Delta_{\rm targ} = 100\times ({\rm HMF} - {\rm HMF_{targ}}) / {\rm HMF_{targ}}$, the percent difference to the real HR halo mass function. The matched HR matches the complete HR to 1 {\rm per~cent} precision. The mass distributions of all and unmatched haloes are also shown for reference. In the insets we show the probability distributions $P(d)$ of a LR halo being a distance $d$ from its match. One can see that the distribution of distance is well smaller than ${\rm r_{match}}$ we use. The peak is at around $0.05 h^{-1} \rm Mpc$}
    \label{fig:matching}
\end{figure}

\reffig{fig:matching} shows the results of the halo matching procedure applied to the train, validation and test sets with $k_{\rm nb} = 10$, $r_{\rm match} = 5~h^{-1}\rm Mpc$, and $\Delta_M^{\rm th} = 0.95$. We show firstly that the halo mass function of the matched HR haloes closely approximates that of the complete HR catalogue. This subset is hereon referred to as Target HR. For clarity, in the sub-panels we show the relative difference between the Target HR and the actual HR HMF, which is 1 {\rm per~cent} in the mass range considered.  We also show the mass functions of all LR haloes and the matched ones. It is evident that the bulk of the unmatched LR haloes is in the low-mass end of the distribution. This is expected, given that the HR mass cut is larger than the analogous cut in the LR catalogue.

The hyperparameters $k_{\rm nb}$ and $r_{\rm match}$ are chosen \textit{ad hoc} to decrease the computational burden of the procedure. The latter permits the kd-tree structure to be pruned, which speeds up the candidate search. The former decreases the number of loops in the mass assignment step, though it is sub-dominant for the overall performance of the halo matching procedure. In fact, $\Delta_M^{\rm th}$ plays the most important role in controlling the HMF of the matched haloes. Moreover, the insets in \reffig{fig:matching} show the probability that a pair of matched halos be separated a distance $d$. Notice that most halos find a match within $d\sim10^{-1}~h^{-1}\,\rm Mpc$. This means that the chosen $r_{\rm match} = 5~h^{-1}\,\rm Mpc$ is large enough to accommodate all matches; and while it allows for faster tree queries, it does not affect the result of the matching itself. \reftab{tab:matching-hyperparams} shows the hyperparameters introduced by the one-to-one halo matching procedure and the values used in this work. The corresponding matching statistics are shown in \reftab{tab:matching-statistics}. Note that the matching statistics are similar for the different sets, which is desired.
\begin{table}
    \centering
    \caption{Hyperparameters for halo matching and their chosen values.}
    \begin{tabular}{@{}lc@{}}
\toprule
\textbf{Parameter}  & \textbf{Value}    \\ \midrule
$k_{\rm nb}$        & 10                \\
$r_{\rm match}$     & $5~h^{-1}\rm Mpc$ \\
$\Delta_M^{\rm th}$ & 0.95              \\ \bottomrule
\end{tabular}
    \label{tab:matching-hyperparams}
\end{table}

\begin{table}
    \centering
    \caption{Matching statistics of the three data sets using the hyperparameter values chosen (see \reftab{tab:matching-hyperparams}).}
    \begin{tabular}{@{}cccc@{}}
\toprule
                       & \textbf{Train} & \textbf{Validation} & \textbf{Test} \\ \midrule
Fraction of HR matched & 0.9983         & 0.9983        & 0.9983        \\
Fraction of LR matched & 0.5414         & 0.5411        & 0.5400        \\ \bottomrule
\end{tabular}
    \label{tab:matching-statistics}
\end{table}

\subsection{Feature selection}

The features our model takes into account play a significant role in its performance. In principle, the FoF halo catalogues provide a number of halo properties (e.g. see \reftab{tab:halo-properties}) that can be used as features of the model. However, these may contain only local information and therefore ignore the structure of the cosmic web. In order to enhance the predictive power of the model, we are interested in including also properties of the environment of each halo.
 
To this end, we attach to each halo the local halo overdensity at its position as $\log\delta_{\rm h}$. The overdensity field is computed in grid sizes $N_{\rm grid}$ of 2048, 1024, 512, 256, 128 and 64 bins along the longest side of the sub-box in order to include information from different scales. These correspond to cubic cell sizes of 0.49, 0.98, 1.95, 3.90, 7.81, 15.63 and 31.25$~h^{-1}\rm Mpc$ respectively. These features are hereon denoted \texttt{log\_dens}$i$ where $i$ indicates the value of $N_{\rm grid}$.

\begin{figure}
    \centering
    \includegraphics[width=\linewidth]{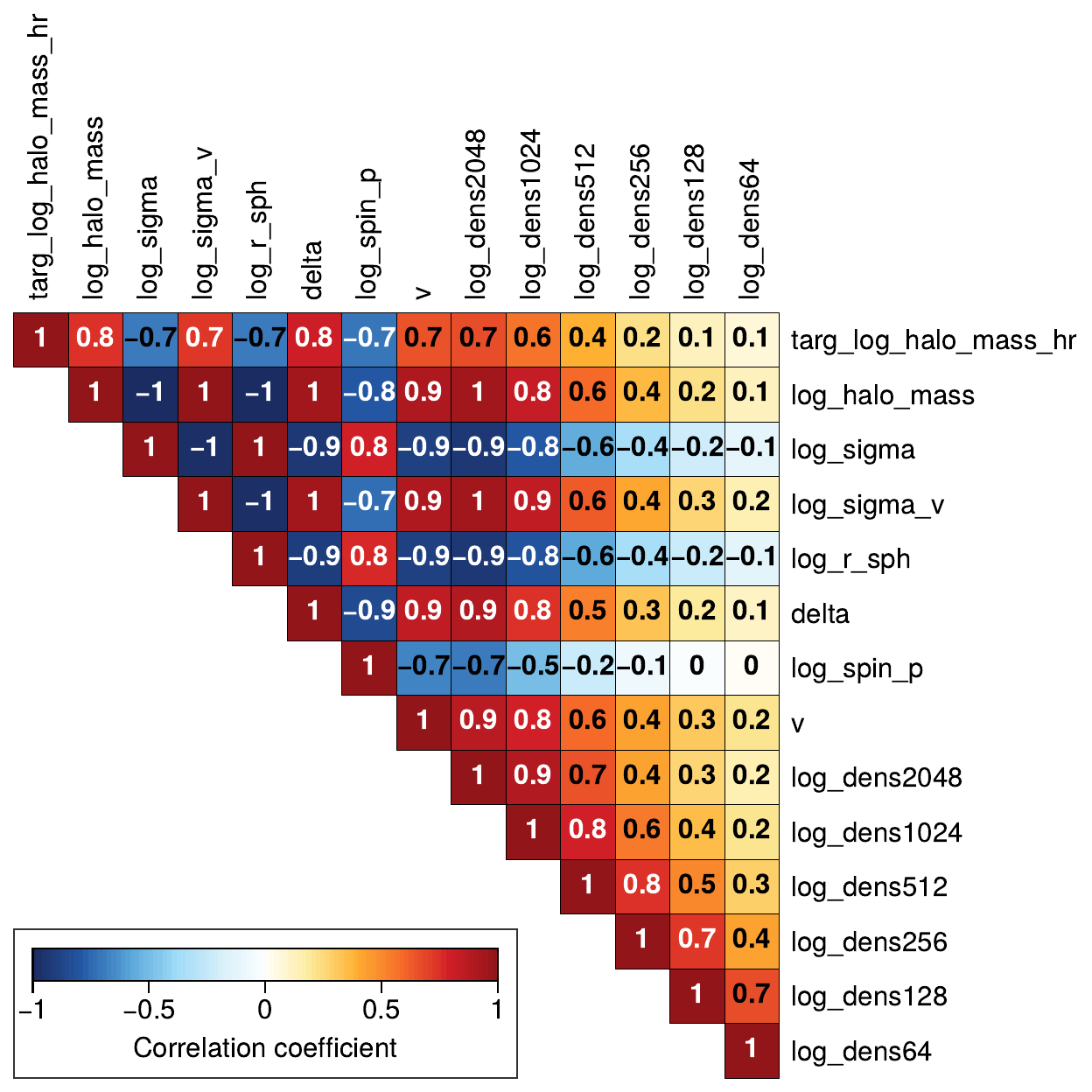}
    \caption{Correlation matrix of the complete set of halo features and the matched HR halo mass (see \refsec{sec:halo-matching}). Colourbar shows the correlation coefficients. Names and descriptions of the features can be found in \reftab{tab:halo-properties} and  \refsec{sec:data}.}
    \label{fig:correlations}
\end{figure}

In \reffig{fig:correlations} we show the correlation matrix of the halo features computed from the test set. Different halo properties are found to be highly correlated among them. These high correlations suggest that these properties provide redundant information to the model and can thus be safely discarded, without degrading the model's performance. Notably, the engineered features, \texttt{log\_dens}$N_{\rm grid}$,  are the ones showing the smallest correlations with the halo properties. Furthermore, the decorrelation with the LR halo mass decreases along the grid size, meaning that averaging the overdensity over a larger neighbourhood does indeed provide more information to the model. Nonetheless, it is not clear if the model will use this extra information efficiently to perform the prediction. We also show the correlations of the features with the target halo mass obtained through the one-to-one halo matching procedure described in \refsec{sec:halo-matching}. The correlation of the HR mass with the LR halo properties is weaker than it is with the LR mass; and may be useful to identify the most relevant features for the model.

\section{Model benchmark}
We test our model against two different techniques to assign halo masses to the LR haloes. 
\subsection{Abundance Matching}
We first test an abundance matching technique, in which we rank halo masses in both the HR and LR catalogues but assign as `prediction' the largest HR masses to the most massive LR haloes and so on. Given that our LR catalogue has more objects that the HR catalogue, the discarded LR haloes in this case are the least massive. This approach, by construction, matches perfectly the HR HMF and ignores any spatial information, so it remains to evaluate how the clustering is affected.

\subsection{\textsc{hadron} code}
The \textsc{hadron} code \citep{Zhao2015} was devised to assign halo masses to approximate mock catalogues, such that the halo clustering in different mass bins matches those of a given reference simulation. This is done by extracting the halo bias from the reference simulation as a multidimensional probability distribution over a set of features. These features are related to the cosmic environment of the halos of given masses, such as whether they are located in voids, sheets, filaments or knots and -- in the case of the latter -- the knot mass. This halo bias distribution therefore encodes not only the local density but also the dynamics of matter flows and halo exclusions. These features are analogously extracted from the mock catalogues such that the bias may be sampled for the required quantity, in this case, the halo mass. The concept of the bias as a multidimensional probability distribution has been shown to be very versatile and powerful. It has allowed not only to assign halo masses but also halo positions in order to create realistic mock catalogues \citep{Balaguera2019, Balaguera2020} and even assign baryonic properties to dark matter catalogues \citep{Sinagiglia2021}.

We stress that the `training' of this code does not involve the one-to-one matched catalogues but relies entirely on the dark matter properties of the HR simulation. For our tests we use the complete $1~h^{-3}\rm Gpc^3$ HR to train and the `test' quarter of the LR. We compute the halo environmental features from a $512^{3}$ dark matter density field. Moreover, we have tuned the \textsc{hadron} parameters, such as the number of bins for probability distributions, to optimize its performance.

\section{Machine Learning Model}
\label{sec:models}
Our main objective can be interpreted as a regression task that computes an appropriate mass for each LR halo. \reftab{tab:matching-statistics} shows that most ($>99 ~{\rm per~cent}$) of the HR haloes are matched, however, the fraction for LR hales is only about 50 {\rm per~cent}. This implies that the model should be able to discard LR haloes that are not matched, resulting in a classification task. In this work the positive and negative classes (i.e. matched and unmatched) are roughly equally sampled, which facilitates training the model for classification; however this is not the case in general. Care should be taken when training a model for classification without taking into account unevenly sampled classes in the training set. 

The input of our model consists in a LR halo object which contains various features. The targets ($\vb*{y}$) of our model are a boolean label indicating whether the halo has a match and a target halo mass (i.e. log-mass). For the haloes that have not been matched (label 0), the target halo mass has been set to zero.

\subsection{Random forests}
The machine learning method we use for this study is a random forest \citep{Ho1995}, which is an ensemble of decision trees. A decision tree is a model that outputs a value or vector based on a set of decision boundaries imposed on the training data. In a geometric sense, the decision tree recursively partitions the feature space. As a result, the decision tree does not infer parameters from the training data, but places test data into these partitions. These trees are known to be sensitive to small changes in the training data. To mitigate this, they are usually implemented in ensembles called Random forests (RFs). The output of a RF is the mode or average of the output of the decision trees, depending on the task the forest is trained for. 

Random Forests are known to be very capable models, but prone to overfitting due to the fact that the trees strongly depend on the training data. A large number of trees ($n_{\rm tree}$) acts as a regularisation for these models, and increasing the maximum depth ($d_{\rm max}$) of the trees increases their capacity. The tree depth is the maximum number of partitions of the feature space per branch, that is, the maximum number of nested if-else conditions imposed on the features. Forests also provide a feature importance metric, which corresponds to the fraction of times each feature is used to set a decision boundary. This can be used to perform feature selection in case some features are found to be of little importance.

In our case, we use the {\sc scikit-learn}'s implementation of a random forest regressor. The random forest outputs two numbers $\vb*{\hat{y}} = (p_{\rm keep}, \log(M_{\rm halo}))$, that correspond to the probability of keeping the halo and the target log-mass, respectively. We implement the classification task as a regression of the probability of keeping the LR halo. We explored the possibility of separating the classification and regression task and applying two separate models in series, however we found that this is not more effective than a single model that performs both tasks in parallel but costs twice as much in terms of memory use.

For this work, the model consists of a random forest with $n_{\rm tree} = 10$ and $d_{\rm max} = 64$. These hyperparameters are fixed to control the memory usage of the model, which strongly depends on $n_{\rm tree}$. We tested on a subset of the training catalogue and with $n_{\rm tree}=100$, which yields comparable results to those of our fiducial choice. The maximum depth is fixed to be a large value due to the large amount of training data (i.e. LR haloes), which means that many partitions are needed to correctly allocate the data.

The loss function used is the mean-squared-error (MSE)
\begin{equation}
    {\rm MSE} = \frac{1}{N_{\rm haloes}} \sum_{n}^{N_{\rm haloes}}(\vb*{y}_n - \vb*{\hat{y}}_n)^2.
\end{equation}
To evaluate the performance of the model we will focus on the halo mass function and the 2- and 3-point clustering as physically meaningful metrics introduced in the next section. Note that these metrics are not included in the loss function.

\section{Evaluation Metrics}
\label{sec:evaluation-metrics}
In this section we introduce the metrics used to evaluate the performance of the model proposed. We first introduce some metrics commonly related to the classification performance in machine learning. Then we introduce the physically relevant metrics such as the halo mass function and two- and three- point clustering statistics.

\subsection{Classification performance metrics}

There are a number of metrics available for evaluating classifiers besides the naive accuracy or fraction of misclassified samples. In a general sense, the relevant quantities that define these metrics are the number of true positives, true negatives, false positives and false negatives, given that these encode the output probabilities conditioned on the true labels. In our study, the true positives are the haloes that were matched and kept (M \& K) by our model, the true negatives correspond to unmatched and discarded (U \& D) haloes, the false positives are the unmatched and kept (U \& K) haloes and false negatives are matched haloes that are wrongly discarded (M \& D). The 2$\times$2 matrix showing the probabilities of the combination of true/false and positive/negative cases, i.e. confusion matrix, will be shown later.

The Receiver Operating Characteristic (ROC) is a curve relating the true positive rate (TPR) 
\begin{equation}
    {\rm TPR}(\gamma) = \frac{\rm TP(\gamma)}{\rm FN(\gamma) + TP(\gamma)}
\end{equation}
to the false positive rate (FPR) 
\begin{equation}
    {\rm FPR}(\gamma) = \frac{\rm FP(\gamma)}{\rm TN(\gamma) + FP(\gamma)}
\end{equation}
and is parametric in the decision threshold ($\gamma$) chosen for the classification. The values $\rm TP$, $\rm FP$, $\rm FN$ and $\rm TN$ denote the number of true positives, false positives, false negatives and true negatives respectively.

We can quantify the performance of the classifier using the Area Under the Curve (AUC) metric, which is defined as the integral of the ROC curve. An AUC value close to 0.5 will therefore imply a diagonal ROC, thus poor performance of the classifier. A perfect classifier would show an AUC close to 1, meaning that the TPR is maximised irrespective of the threshold $\gamma$.

\subsection{Halo mass function}

The Halo Mass Function (HMF) used through this work is defined as the distribution of halo masses. This statistic is of physical relevance due to the fact that the mass distribution of dark matter haloes is closely related to the evolution of the Universe, especially in the matter domination era \citep{Lukic2007}. However, the accuracy on the lower mass end is not critical for constructing galaxy catalogues since we do not assign galaxies to every haloes with lower halo mass \citep{Alam2020}. In this case, an unbiased mass-clustering relation (respecting to HR) is more useful for constructing a robust galaxy catalogue. We introduce the clustering statistics in the following sections.

\subsection{Power spectrum and Two-point correlation function}

The two-point information encoded in the large-scale structure of the Universe is also of utmost importance for cosmology \citep{Alam2017, Alam2021}. We therefore test the two-point clustering of our model's output in Fourier and configuration space. In the former, the relevant statistic is the power spectrum $P(k)$, defined as
\begin{equation}
    \label{eq:power-spectrum}
    (2\uppi)^3\delta^{D}(\vb*{k}-\vb*{k'})P(k) = \langle\delta(\vb*{k})\delta(\vb*{k'})\rangle,
\end{equation}
where $\delta^{D}$ is a Dirac delta distribution, $\vb*{k}$ and $\vb*{k'}$ are wavevectors, $\delta(\vb*{k})$ is the Fourier-space overdensity field and $\langle\,\cdot\,\rangle$ denotes averaging. Through this work we estimate power spectra using the \textsc{Nbodykit} implementation \citep{Hand2019}. 

The peculiar velocity of observed objects influences the redshift measurements such that our observations are in redshift space, where the observed tracer distribution is not isotropic. Under the plane-parallel approximation, we compute the redshift space distortions and add them in the real-space cartesian $Z_{r}$ direction:
\begin{equation}
    \label{eq:rsd}
    Z_{z} = Z_r + \frac{v_{\rm pec}}{aH}.
\end{equation}

$Z_z$ denotes the redshift space position along the third axis, $v_{\rm pec}$ is the halo's peculiar velocity along the line of sight, $a$ is the cosmological scale factor and $H$ the Hubble parameter.

The anisotropy induced by the redshift space distortions can be detected in the multipoles of the two-point correlations. Multipoles $P_\ell(k)$ are computed from the 2-dimensional power spectrum $P(k,\mu)$, where $\mu$ is the cosine of the angle to the line of sight, they are defined by
\begin{equation}
    \label{eq:power-multipoles}
    P_\ell(k) = \frac{2\ell+1}{2}\int_{-1}^1\dd\mu~P(k, \mu)L_\ell(\mu),
\end{equation}
where $L_\ell(\mu)$ is the $\ell$-th Legendre polynomial.

The two-point correlation function $\xi(s)$ is the Fourier pair of the power spectrum. We estimate it using the natural estimator \citep{Pebbles1974}:
\begin{equation}
    \label{eq:tpcf}
    \xi(s,\mu) = \frac{{\rm DD}(s,\mu) - {\rm RR}(s,\mu)}{{\rm RR}(s,\mu)}.
\end{equation}
The term $DD(s,\mu)$ stands for the number of pairs separated by a distance $s$ in the data catalogue, normalised by the total number of pairs $N_D(N_D - 1)$ where $N_D$ is the number of objects in the catalogue. Similarly the $RR(s,\mu)$ terms is the number of such pairs in a random catalogue. For our periodic simulation data, the $RR$ factor is computed analytically.

Analogous to the power spectrum, the multipoles of the correlation function are defined as
\begin{equation}
    \label{eq:tpcf-multipoles}
    \xi_\ell(s) = \frac{2\ell+1}{2}\int_{-1}^1\dd\mu~\xi(s, \mu)L_\ell(\mu).
\end{equation}

\subsection{Bispectrum}

Similar to the power spectrum, the bispectrum $B(k_1, k_2, k_3)$ is the Fourier pair of the three-point correlation function. It is defined as 
\begin{equation}
    \label{eq:bispectrum}
    \delta^{D}(\vb*{k}_1+\vb*{k}_2+\vb*{k}_3)B(k_1, k_2, k_3) = \langle\delta(\vb*{k}_1)\delta(\vb*{k}_2)\delta(\vb*{k}_3)\rangle.
\end{equation}
In this work we evaluate the bispectrum in a configuration with $k_1 = 0.2$ and $k_2 = 0.4~h\,\rm Mpc ^{-1}$. $k_3$ is defined for different angles $\theta$ by $k_3^{2} = k_2^2\sin^2\theta + (k_2\cos\theta + k_1)^2$ such that the wavenumbers form triangles of sides $k_1$, $k_2$, $k_3$. 

The bispectrum contains important information on the large scale structure of the distribution of tracers and can be used to constrain the tracer bias. It has also been used to constrain various cosmological parameters \citep{Sefusatti2006, Gil2017}. Moreover, due to its nonlinear nature, it is a relevant statistic to constrain primordial non-Gaussianity \citep{Gangui1994, Sefusatti2007, Welling2016} and deviation from General Relativity \citep{Borisov2009, Gil2011}. Because of this, a precise estimation of the bispectrum is desired from the model.

In this work we use the \textsc{Pylians 3}\footnote{\url{https://pylians3.readthedocs.io/en/master/index.html}} library to compute the bispectra.

\section{Results}
\label{sec:results}
In this section we present the results of the random forest model used on the test set. We experiment with a varying number of features among the ones shown in \reffig{fig:correlations} and find that the number of features included is an important hyperparameter. We observe that a large number of features allows the model to closely fit the HMF but has a detrimental effect on the power spectra by introducing a negative bias with respect to the HR mass-binned spectra. In what follows we show the results that correspond to the use of three features. These are chosen to be the most positively correlated ones with the target halo mass (\texttt{log\_halo\_mass\_hr}, \reffig{fig:correlations}): \texttt{log\_halo\_mass}, \texttt{log\_sigma\_v} and \texttt{delta}. In additional to the model with three features, we will also show a model with four features (i.e., add one more feature: \texttt{log\_sigma}) for comparison.

\subsection{Halo abundance}
\begin{figure}
    \centering
    \includegraphics[width=\linewidth]{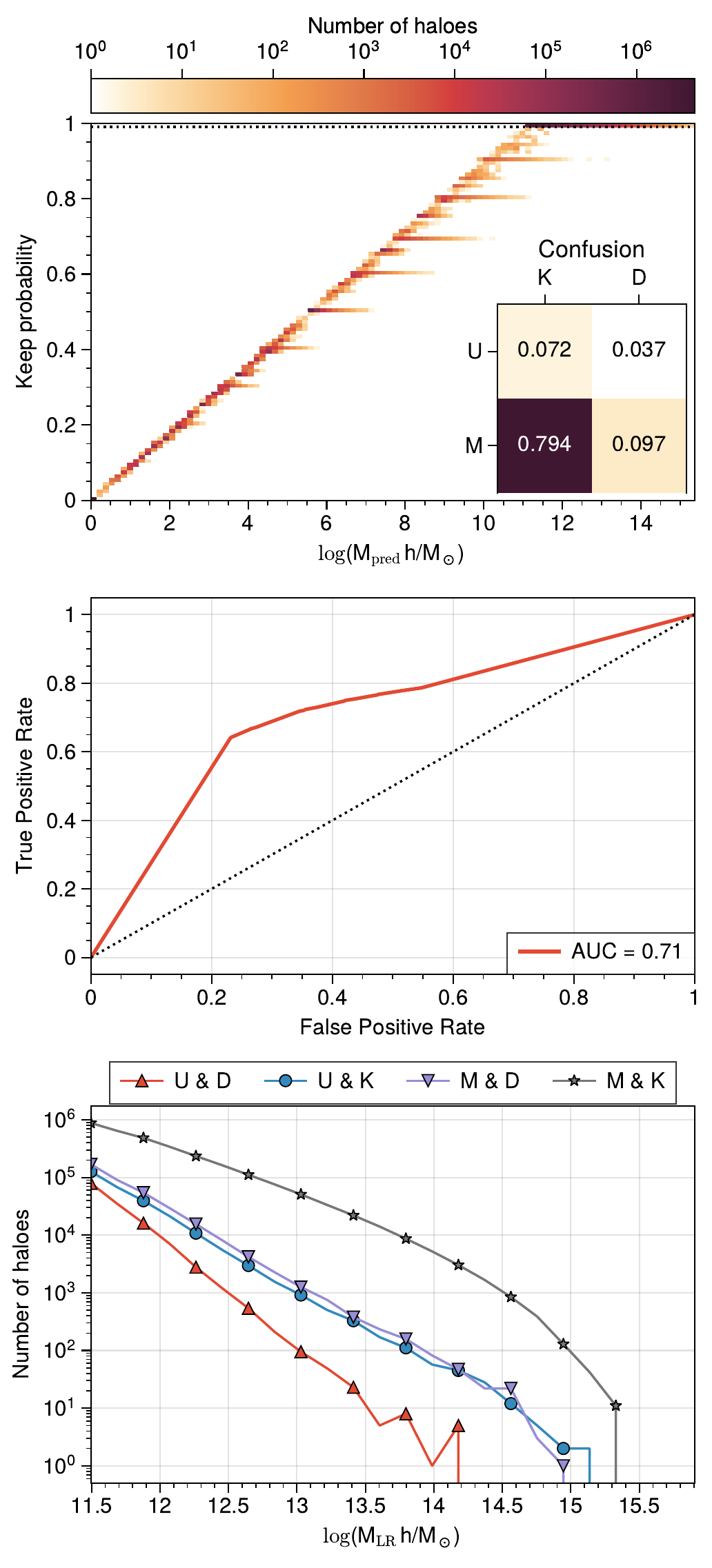}
    \caption{\textit{Top}: Probability of keeping the halo $p_{r\rm keep}$ as a function of the predicted halo mass $M_{\rm pred}$. Horizontal dotted line shows the threshold keep probability of 0.99.
    \textit{Middle}: ROC to evaluate the classification performance of the model. The diagonal shows the expected curve from a random classifier.
    \textit{Bottom}: Halo mass distributions of haloes under different conditions. `U': Unmatched, `M': Matched, `K': Kept, `D': Discarded. In this sense, U\&K denotes false positives, U\&D true negatives, M\&D false negatives and M\&K true positives.}
    \label{fig:discard}
\end{figure}

The first task our model should perform is to be able to discard some haloes in the input to account for the LR haloes that were not matched. The true labels for this task are therefore matched (`M') and unmatched (`U'), while the predicted labels are either kept (`K') or discarded (`D'). 

\reffig{fig:discard} shows the results of this task. The keep probability $p_{\rm keep}$ is shown to be highly correlated to the predicted halo mass $\log (M_{\rm pred})$, as is expected. Our model does predict halo masses continuously between zero and the maximum training halo mass. We must then choose a threshold value for $p_{\rm keep}$ such that only relevant masses are included. It is also seen that the vast majority of haloes are predicted to be in the relevant mass range ($\log (M_{\rm pred}~h/M_\odot) > 11.4$). We select the threshold to be at 0.99, which means that all the decision trees must have had a high output confidence of the halo being matched. 

We use the ROC to show that the classifier has indeed learnt to identify which haloes should be removed, which is better quantified in the value of the AUC of $0.71$. In addition, from the ROC it is clear that aiming for a high TPR -- as we have done -- comes at the cost of increasing the FPR. We suspect the performance of this classifier would improve if it were trained using a loss function more suitable for classification than the MSE.
\begin{figure}
    \centering
    \includegraphics[width=0.99\linewidth]{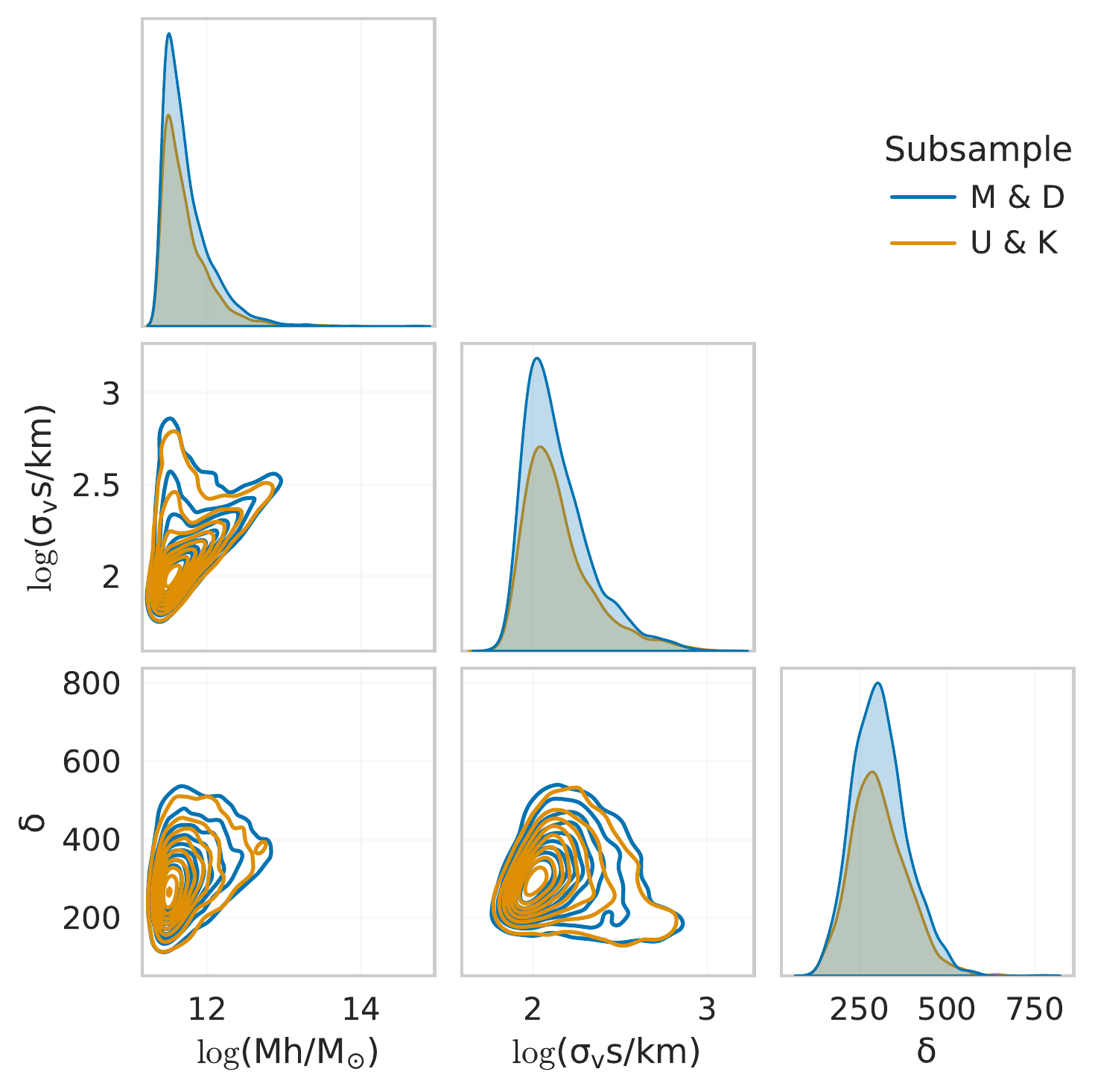}
    \caption{Corner plot showing the distribution of properties of a subset of haloes from the false positives (U \& K) and false negatives (M \& D) subsamples. The halo properties of these are very similar, thus validating the model performance despite the classification errors.}
    \label{fig:match-environments}
\end{figure}

We then estimate the impact of a high FPR in the physically motivated evaluation metrics. Firstly, we observe the confusion matrix of the predictions (inset in \reffig{fig:discard}) which show that the fraction of false positives (incorrectly kept haloes) and false negatives (incorrectly discarded haloes) are similar (~8 {\rm per~cent}). If the haloes in these subsets have similar masses and are located in similar environments, it would be reasonable to assume that the samples are roughly interchangeable. We plot the halo mass functions of these subsamples (bottom panel \reffig{fig:discard}) and observe that their mass distributions are consistent with one another. Given the similar fractions of FP and FN, we further assume that if an unmatched halo is misclassified, it would most likely be replaced by a similar halo which locates at a similar environment (see \reffig{fig:match-environments}).
This would imply that interchanging haloes among the FP and FN samples does not introduce very significant biases in the HMF and clustering measurements. Because of this, we keep the classifier trained with the MSE loss and leave a more complex training scheme for future work.

Finally, we find that in the mass range of $10^{11.4}$ to $10^{15.5}~h^{-1}M_\odot$, the classifier keeps over 80 {\rm per~cent} of the samples, which is in line with the fact that the largest number of unmatched LR haloes was located in the lowest mass end, where the LR and HR HMF do not overlap  (log mass $< 11.4$), while for larger masses, most LR haloes are matched (see \reffig{fig:matching}). 

\subsection{Halo mass function}
\begin{figure}
    \centering
    \includegraphics[width=\linewidth]{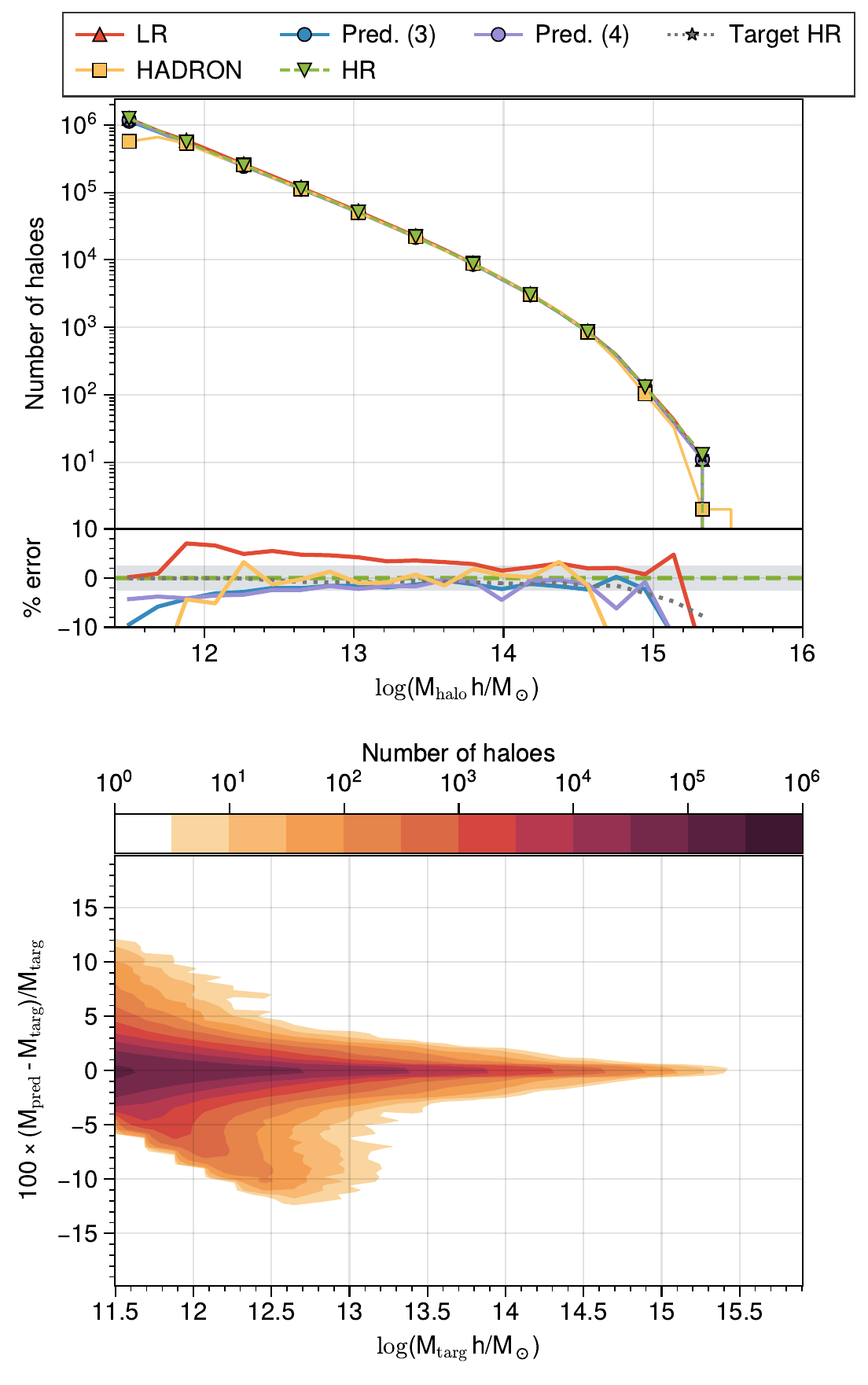}
    \caption{Top:  Halo mass functions of the low resolution, high resolution, prediction and target catalogues. We show the prediction of models with three (Pred. (3)) and four (Pred. (4)) features. The subpanel shows the percent difference of the LR, Target HR and Prediction halo mass functions with respect to the real HR HMF. Shaded area highlights the 2.5 {\rm per~cent} error band. Bottom: Map showing the distribution of the percent error of the predicted masses against the target masses as a function of the target mass.}
    \label{fig:hmf}
\end{figure}
Through this subsection we evaluate the performance of the model in terms of purely predicting the HR halo masses, i.e. the regression task. We compare with a second model that uses an extra feature: \texttt{log\_sigma}. These two models show the best overall performance. We find that naively including all halo features and engineered quantities does yield a good performance in terms of HMF but is detrimental to the clustering statistics (see \refsec{sec:2pt}).

In \reffig{fig:hmf} we plot the halo mass functions of the LR catalogue, the HR catalogue, the Target HR (the subset of HR that was matched), the \textsc{hadron} prediction and the predicted HR from the models with tree and four features. The agreement between all of these is good in this mass range. The sub-panel shows the percent error difference of the different samples with respect to the real HR. The predicted halo mass function from the model with 4 features closely approximates both the target and the real HR HMF. The prediction using three features shows a slightly larger discrepancy with the reference, however the agreement is still better than 3 {\rm per~cent} from $\log (M/M_\odot) \sim 12.5$ up to $\log (M/M_\odot) \sim 14.5$. The \textsc{hadron} prediction has a similar behaviour to the RF for low masses, but deviates from the reference at a smaller mass on the large-mass end.

Moreover, we compare the distribution of target vs predicted values from our fiducial (3 feature) model using the relative error to the target. We note that the majority of the haloes ($\sim$ $90$ {\rm per~cent}) lie well within the 5 {\rm per~cent} error in the prediction, while a small percentage of the catalogue shows errors of up to 12 {\rm per~cent}. Notably, for the majority of the haloes the error in the prediction increases as the mass decreases.

\citet{Cui2012} have shown that the presence of baryons can bias the HMF by a few per cent with respect to a dark matter-only simulation. Moreover \citet{Beltz2020} have shown that this bias can be as large as 20 per~cent for low masses and that differences between the HMF with and without baryons can be highly dependent on the hydrodynamical model used in the simulations. While we have not explicitly included baryonic effects in our model, our predictions are accurate enough for Halo Occupation Distribution (HOD) modelling. Note that, in practice, a few percent error in the HMF at lower mass region is not be critical for building a galaxy catalogue using the galaxy-halo relation such as HOD model, since the model will randomly drop a significant fraction of haloes anyway. Thus, for modelling galaxy clustering, having correct halo clustering statistics is more important than an accurate HMF. 

A similar model to the one presented in this work can potentially be trained to learn to include baryonic effects in the output or even predict some baryonic properties. We reserve these studies for future work.

\subsection{Two-point clustering}
\label{sec:2pt}

As mentioned, for galaxy clustering modelling, we are interested in whether the model is able to reproduce large scale structure statistics in the HR data. In this section we explore the two-point statistics of the predictions which are not explicitly optimized by our model.

\begin{figure}
    \centering
    \includegraphics[width=\linewidth]{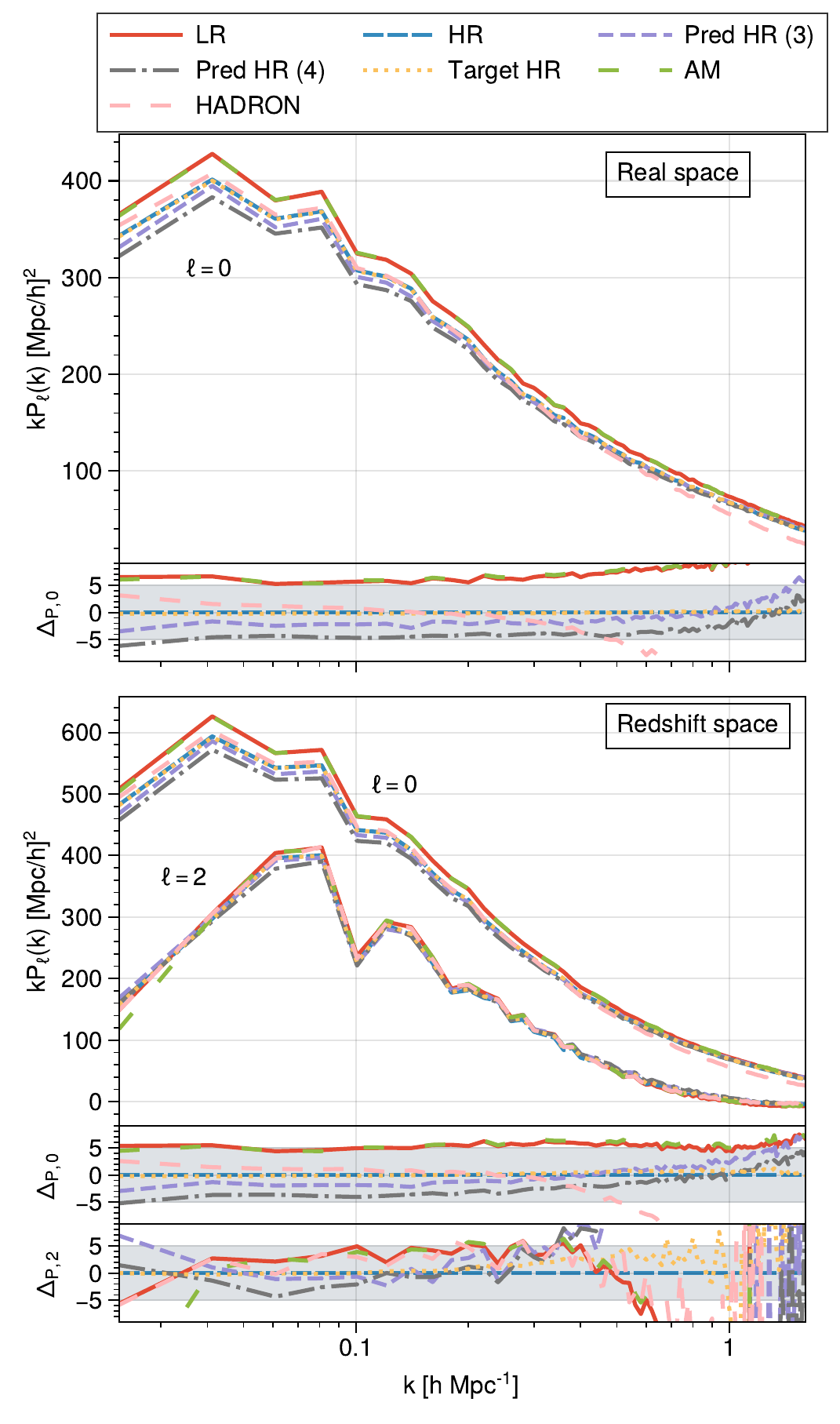}
    \caption{Power spectra monopoles and quadrupoles in real (top panel) and redshift (bottom panel) space. The sub-panel in the top shows the percent error to the HR power spectrum. The sub-panels in the bottom show the percent error defined as $\Delta_{P,\ell} = 100\times\frac{P_\ell - P_{\ell,\rm HR}}{P_{\ell,\rm HR}}$ in both the mono- and quadrupoles. The shaded area highlights the 5 {\rm per~cent} error range.}
    \label{fig:pk-complete}
\end{figure}

\begin{figure}
    \centering
    \includegraphics[width=\linewidth]{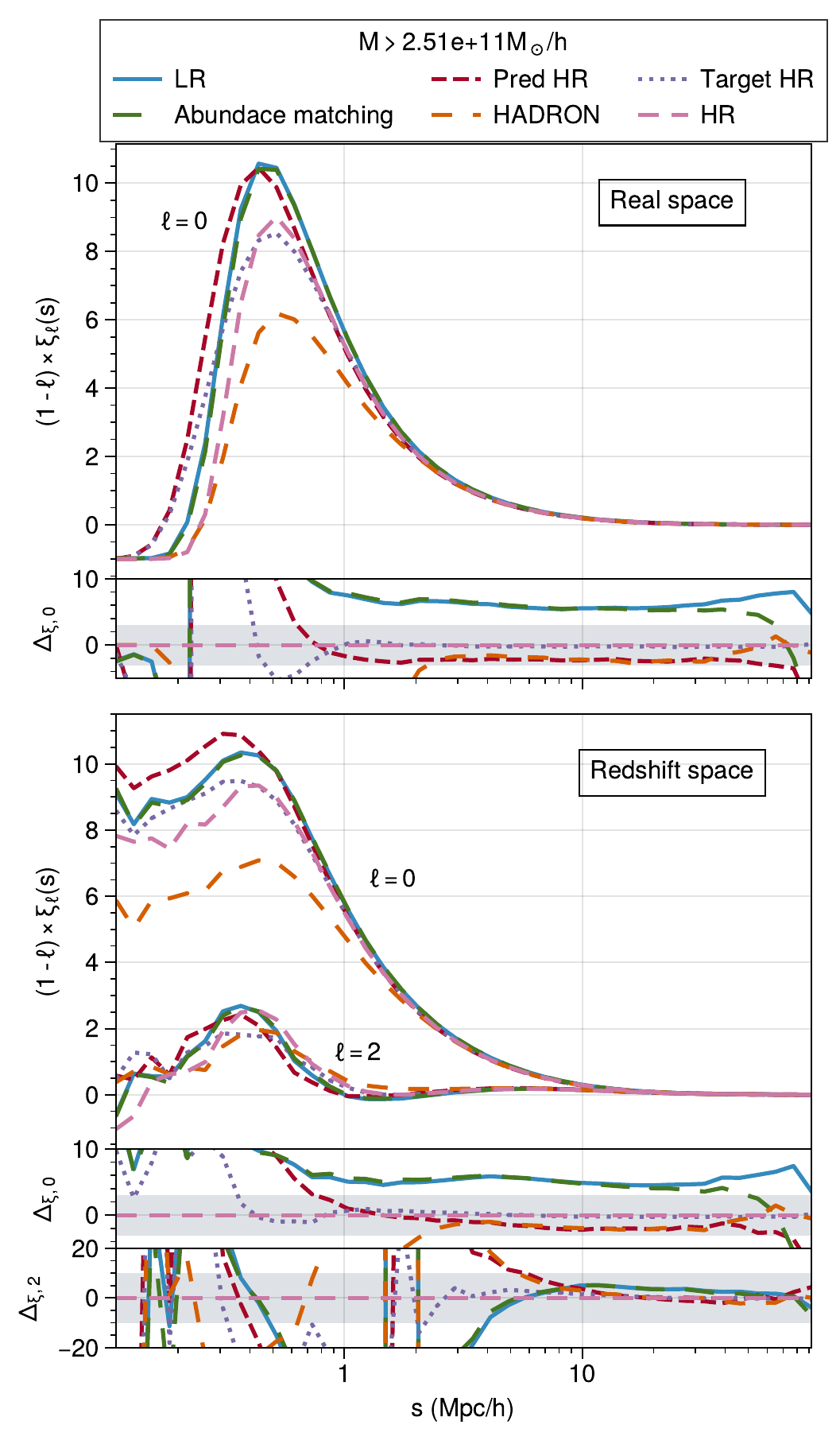}
    \caption{Correlation function monopoles and (negative) quadrupoles in real (top panel) and redshift (bottom panel) space. The sub-panel in the top shows the percent error to the HR correlation function. The sub-panels in the bottom show the percent error defined as $\Delta_{\xi,\ell} = 100\times\frac{\xi_\ell - \xi_{\ell,\rm HR}}{\xi_{\ell,\rm HR}}$ in both the mono- (top) and quadrupoles (bottom). The shaded area highlights the 3 (10) {\rm per~cent} error range for the monopole (quadupole).}
    \label{fig:xi-complete}
\end{figure}

\reffig{fig:pk-complete} shows the power spectra multipoles of the input/output data as well as the predictions of the abundance matching approach and our RF model. In order to compute the power spectra in redshift space, we apply RSD using the UNIT cosmology. The velocities used for the predicted catalogues are the ones originally in the LR catalogue.

The abundance-matched two-point correlation does not differ significantly from the LR power spectrum in neither real nor redshift space. In contrast, the clustering of the catalogue predicted with our 3-feature RF does match the HR power spectra within a 3 {\rm per~cent} error in the for $k\in(0.03, 1)~h\,\rm Mpc^{-1}$. The 4-feature RF shows a larger negative bias with respect to the reference but is still well within the 5 {\rm per~cent} error band. We remark that though the velocities are not modified by our model, the resulting redshift space clustering is consistent with that of the actual HR catalogue. In fact, the power spectrum quadrupole of the prediction closely match (<5 {\rm per~cent} error) the HR power spectra up to $k\sim0.4~h\,\rm Mpc^{-1}$.

The fact that our multiple-feature model perform better than the abundance matching model shows that information other than LR halo mass is necessary for recovering the clustering statistics of a HR halo sample.
Through the halo one-to-one matching procedure we encode highly relevant environmental information such that the model recovers the reference clustering accurately. The engineered features added to the data were in the end not as relevant. We performed tests with higher numbers of features and find that some of the environmental densities are used by the forest, though with a low relative importance.

Moreover, the power spectra from \textsc{hadron} are also within the 5 per cent error range up to $k\sim 5~h\,\rm Mpc^{-1}$. This implies that the environmental features encoded in the halo bias relation extracted by \textsc{hadron} are useful in recovering the clustering statistics on large scales. On small scales, however, the \textsc{hadron} results do show a large negative bias of the power spectrum. This may be the result of the random downsampling the code is forced to do in order to match the number of objects from the reference. It is possible a modified version of \textsc{hadron} would perform better for this precise task. We explore this possibility in a future study. 

We also evaluate the clustering in configuration space in \reffig{fig:xi-complete}. In this figure we show only the prediction of our fiducial three-feature model and the abundance matching approach. The monopole error over separations of $0.5~h^{-1}\rm Mpc$ is consistent with the HR reference within 3 {\rm per~cent}. However it should be noticed that the target HR also shows a significant deviation from the complete HR catalogue on the smallest scales. We attribute these differences to the matching procedure, given that small scales would be most sensitive to the procedure. Furthermore, the deviations from the complete HR reference of the target and our model's prediction are consistent with each other. To further improve the performance, besides calibrating halo masses, we will need to calibrate the halo positions, we leave this for a future study. 

Our model is also able to improve upon \textsc{hadron}'s results, which deviate over 2.5 per cent from the reference at scales of $\sim2~h^{-1}\,\rm Mpc$ in both real and redshift space. These deviations are also consistent with the larger error on large $k$ observed in the Fourier space results.

Despite the deviations due to matching, the improvement of our model over the LR catalogue is significant. The redshift-space 2PCF quadrupole seems more sensitive to the matching, given that the target shows deviations from the HR catalogue for $\sim 1~h^{-1}\rm Mpc$. Our model's prediction also does not seem to perform as well in this metric and shows deviations of over 20 {\rm per~cent} at these scales.

\begin{figure*}
    \centering
    \includegraphics[width=\linewidth]{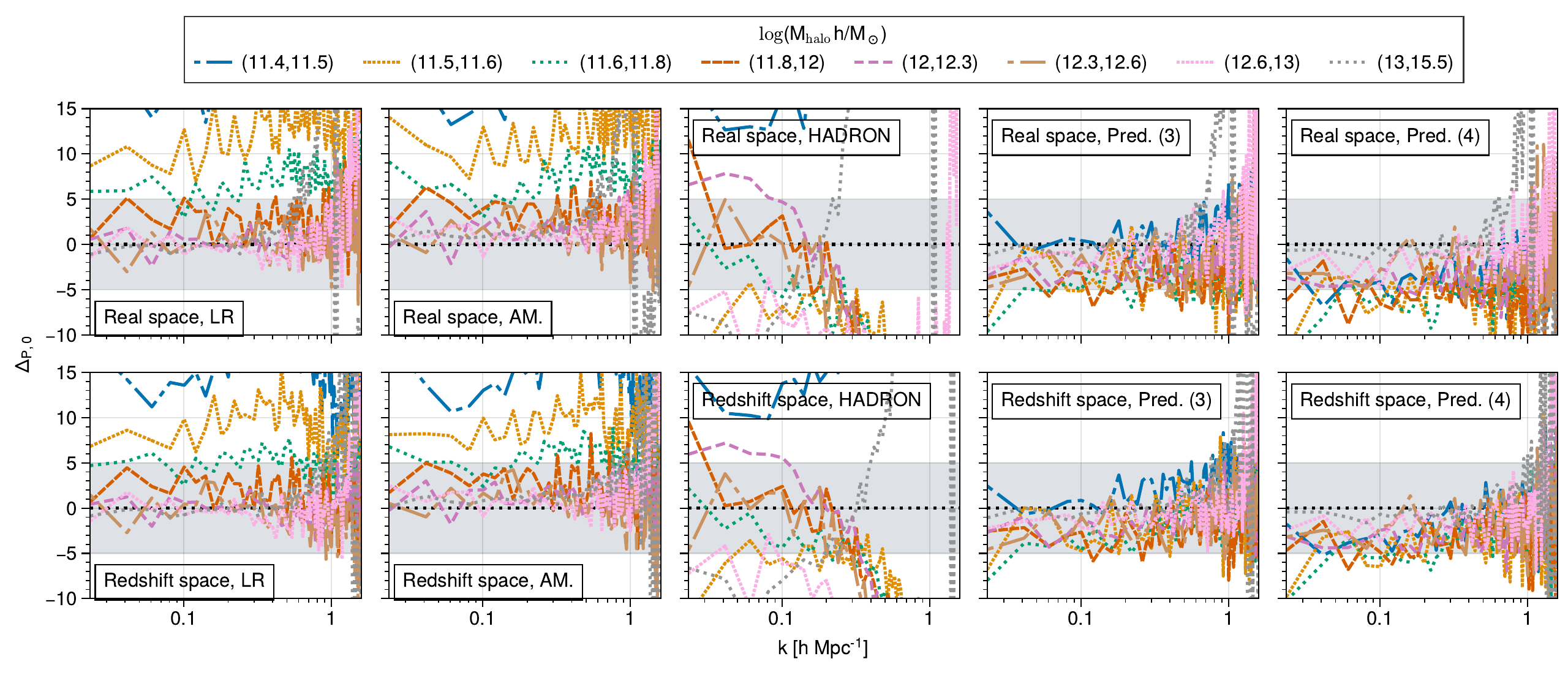}
    \caption{Percent error to the mass-binned HR power spectrum. The leftmost panels show the ratios of the LR power spectra, the middle panels correspond to the abundance matching (AM) approach and the rightmost panels show the results of our RF model. The shaded area highlights the 5 {\rm per~cent} error range. Our model succeeds in matching the HR power spectra to 5 {\rm per~cent} accuracy in the mass bins shown.}
    \label{fig:pk-bins}
\end{figure*}

\begin{figure*}
    \centering
    \includegraphics[width=\linewidth]{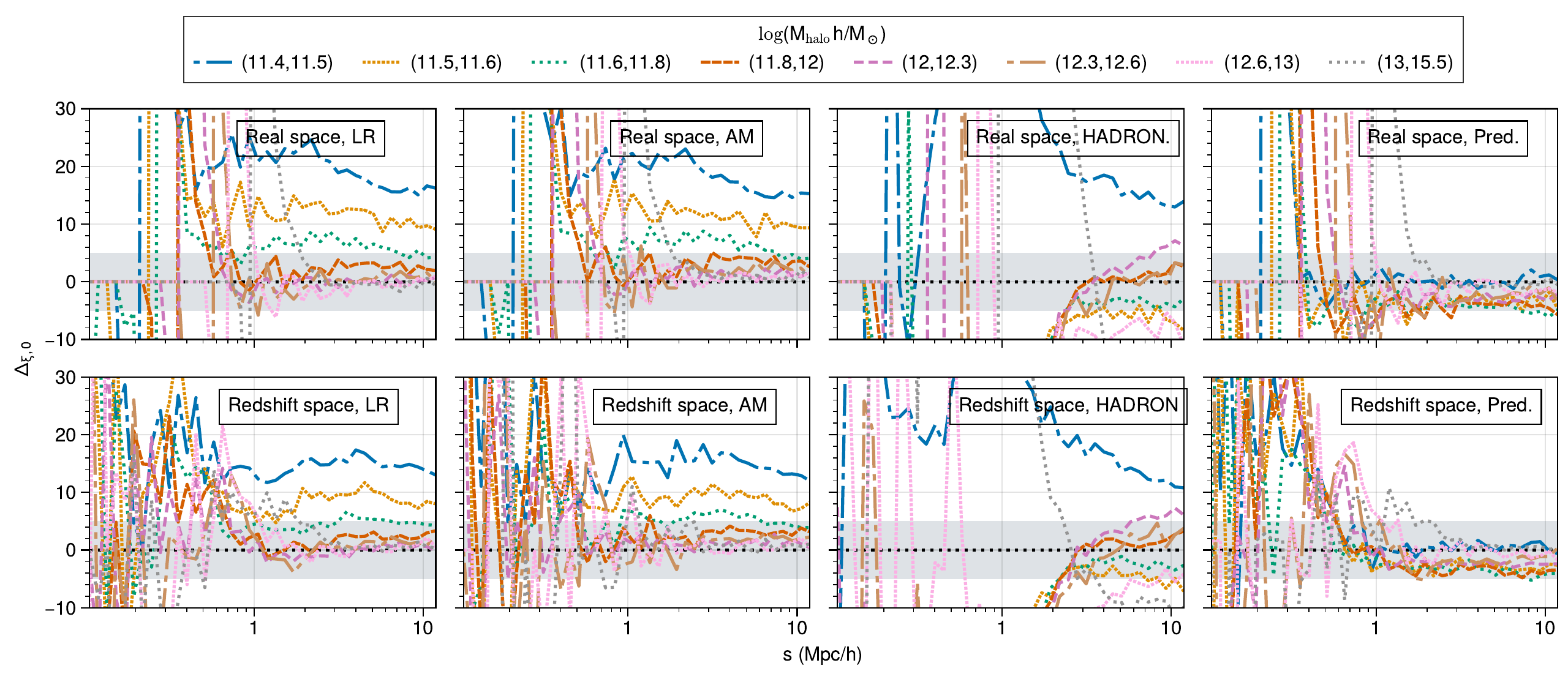}
    \caption{Percent error to the mass-binned HR correlation function. Analogous to \reffig{fig:pk-bins}. Configuration space results confirm that our model succeeds in matching the HR 2PCFs to 5 {\rm per~cent} accuracy in the mass bins shown in scales larger than $1~h^{-1}\rm Mpc$. Large deviations near $s\sim0.3~h^{-1}\rm Mpc$ are due to the 2PCFs crossing zero.}
    \label{fig:xi-bins}
\end{figure*}

In addition, we are interested in recovering the clustering in mass bins. \reffig{fig:pk-bins} shows the percent error difference of the mass-binned power spectra relative to the analogous HR spectra. Once more we observe that the abundance matching coincides with the LR results and is not effective in recovering the clustering. Moreover \textsc{hadron} does influence the mass-binned clustering significantly and yields an error of less than 10 per cent up to $k\sim0.5~h\,\rm Mpc^{-1}$ for all but the least massive bin. This may be due to the fact that \textsc{hadron} prioritises more massive halos when implementing halo exclusions. In contrast, our 3-feature RF (center-right panels) is able to match the target with under 5 {\rm per~cent} error for all mass bins. Evidently, the low mass bins are the ones that show the largest improvement, while for the large mass bins our model introduces a small negative bias. When using the 4-feature RF (rightmost panel) the bias introduced is larger; specially in the low mass bins which seem to be over corrected with respect to the reference. Moreover, we experimented with larger numbers of features and found that, while the HMF is more closely recovered (see \reffig{fig:hmf}), the bias introduced in the clustering was even larger.  There is a trade-off between approximating the HMF and the two point clustering. In choosing a fiducial model, we take into account that a precise clustering is more important than a precise low mass HMF when constructing galaxy catalogues that mimic observations. We therefore chose our fiducial 3-feature model such that the HMF is sufficiently well approximated and the bias in the clustering is small enough.

We further test the massed binned clustering in configuration space in \reffig{fig:xi-bins}, where we no longer show the four feature model results. The configuration space results confirm the Fourier space observations, with 2PCFs resulting from the RF model consistent with the HR clustering at 5 {\rm per~cent} level for $s > 1~h^{-1}\rm Mpc$. Smaller scales show large discrepancies consistent with the ones observed in \reffig{fig:xi-complete}. We observe that larger mass bins start showing a large discrepancies at larger scales than the lower mass counterparts. The reason is that larger mass haloes suffer exclusive effect at larger scales which make the correlation function to be zero or even negative. We would see large percentage error while the correlation function is close to zero.

\subsection{Three-point clustering}

Higher order statistics are also useful for extracting cosmological constraints and studies galaxy properties.
Therefore, we also compare the three-point clustering of our model's output against the HR reference. We use a configuration with $k_2=2k_1=0.4~h\,\rm Mpc^{-1}$. 

\begin{figure}
    \centering
    \includegraphics[width=\linewidth]{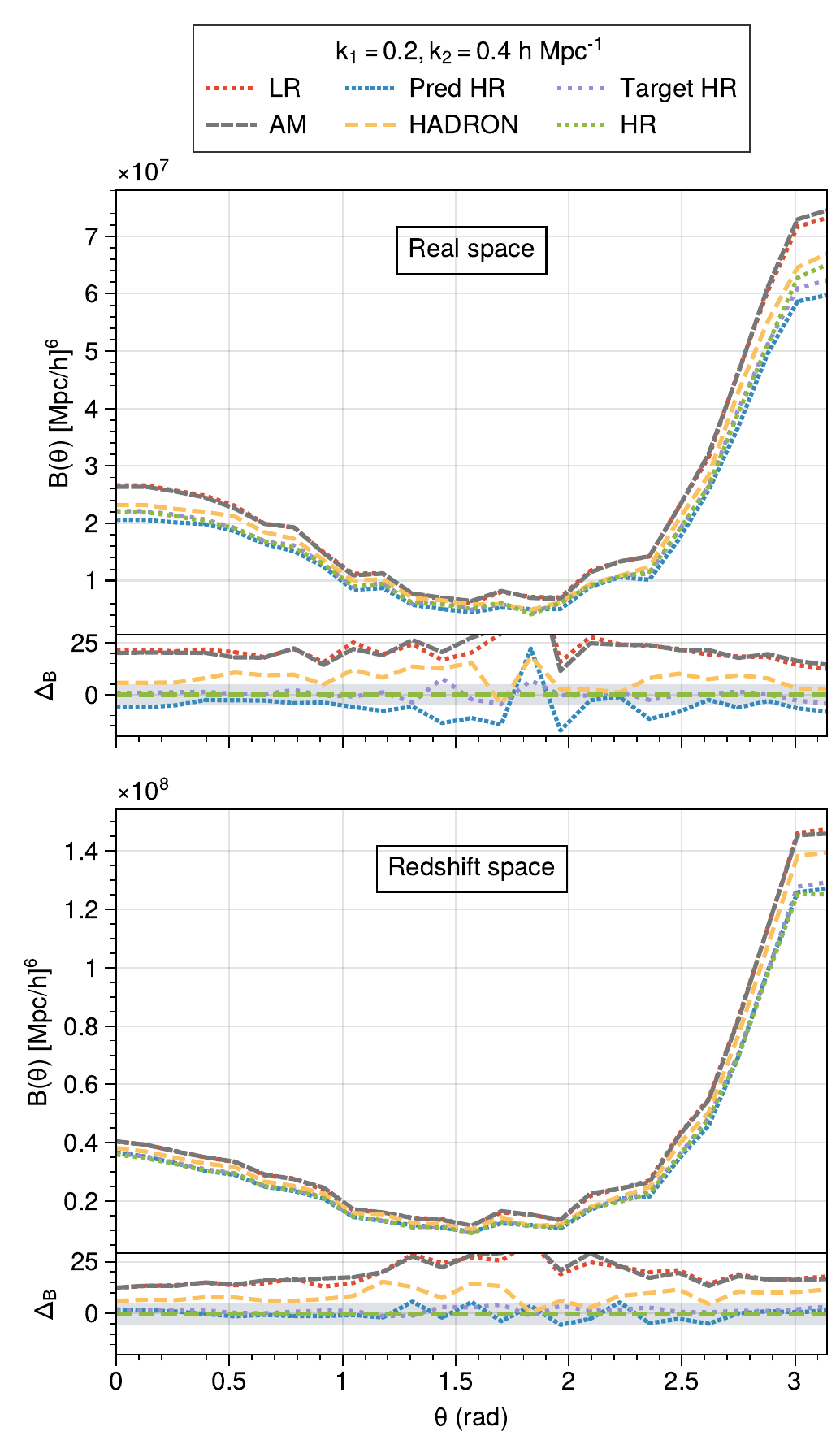}
    \caption{Analogous plot to \reffig{fig:pk-complete} for the reduced bispectra of the different catalogues. The percent error difference is defined as $\Delta_B = 100\times\frac{B - B_{\rm target}}{B_{\rm target}}$.}
    \label{fig:bk-complete}
\end{figure}

\reffig{fig:bk-complete} shows the real- and redshift- space bispectra for the different samples and their percent difference to the HR reference. In real- and redshift- space, the LR and abundance matching curves overlap, which confirms that the mass information is not enough to correct the LR catalogue masses. In addition, these curves show a percent error of $\sim20-25 {\rm per~cent}$ over the whole $\theta$ range. Moreover, we note that the Target HR curve differs slightly ($\sim 5 {\rm per~cent}$) from the real HR bispectrum as $\theta$ approaches $\pi$, however the curves approximate each other from $\theta=0$ up to $\theta=3$. \textsc{hadron} is able to retrieve bispectra consistent with the reference within a $\sim 15$ per~cent difference in both real ad redshift space. Our three feature prediction shows a significant improvement over the LR bispectra, specially in redshift space where the predicted bispectrum lies within a 5 {\rm per~cent} error to the reference HR. In real space the bispectrum of our prediction is slightly negatively biased by $\sim5  {\rm per~cent}$ but is noisy near $\pi/2$. Overall, our prediction shows a 20 {\rm per~cent} improvement in the bias of the bispectra with respect to the reference.

\begin{figure*}
    \centering
    \includegraphics[width=\linewidth]{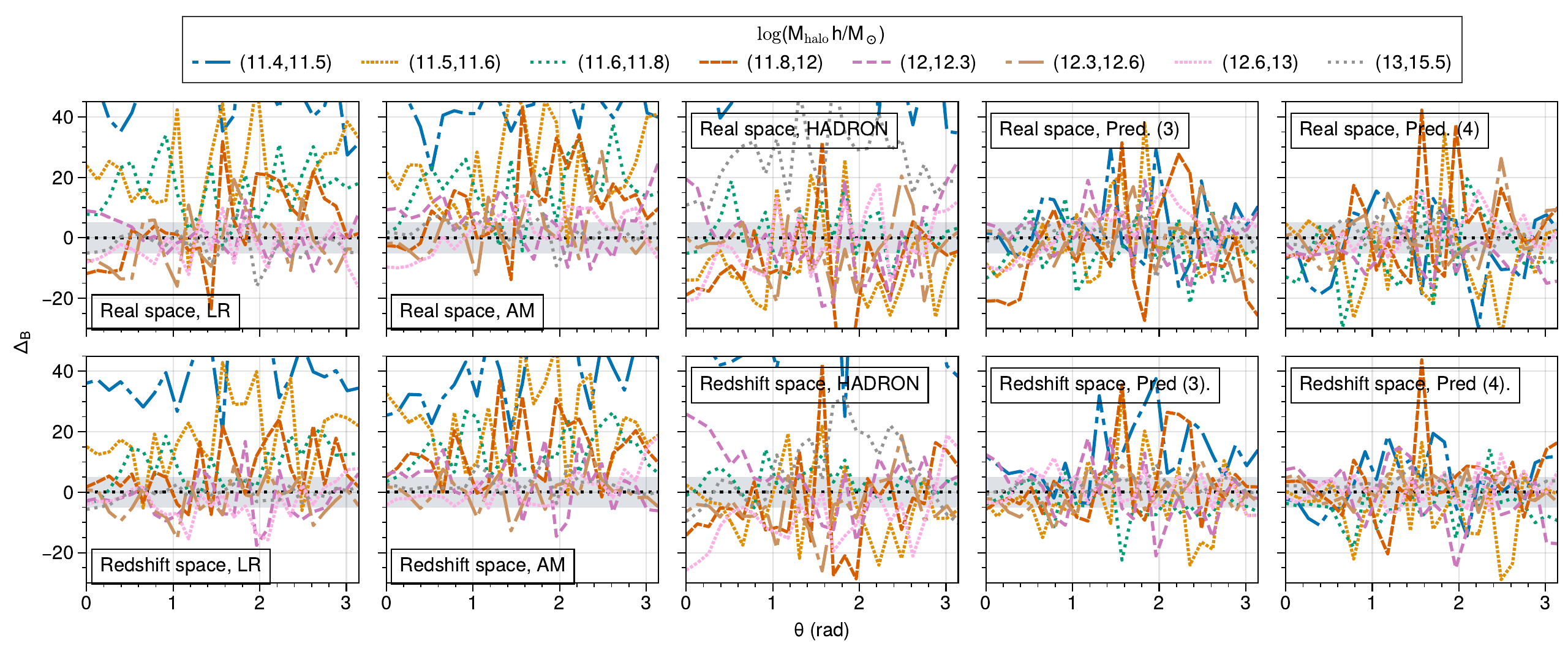}
    \caption{Percent errors in the mass-binned three point statistics with respect to the HR sample. Analogous to \reffig{fig:pk-bins}. We show the 5 {\rm per~cent} error region for reference.}
    \label{fig:bk-bins}
\end{figure*}

Analogous to our analysis of the two point information, we compute the bispectra in mass bins and report the percent differences to the reference. \reffig{fig:bk-bins} shows these results. We observe again that the abundance matching fails to correct the bias of the low-mass tracers and may even bias the high mass tracers, especially in redshift space. Despite the relatively good performance of \textsc{hadron} over the whole mass range, the mass binned bispectra are often more biased than the LR ones. \textsc{hadron} helps in recovering the bispectra for some mass bins, but introduces larger biases for others, such as the most and least massive ones. This is possibly because \textsc{hadron} only implements a single halo exclusion length for all haloes. A mass-dependent exclusion distance may help solve this issue. We explore this in future work. On the fourth panel of the same figure, we observe that the prediction of our 3-feature model shows three point clustering that is consistent with the reference in all the mass bins shown. Even though the curves obtained show an oscillating behaviour, they seem to oscillate around zero. For completeness, we add the results of the 4-feature model in which we observe similar results to the 3-feature bispectra for most mass bins. The least massive bin from the 4-feature model is less biased than the 3-feature, which is consistent with the larger negative biased observed in the two-point clustering of the 4-feature compared to the 3-feature model. 

\section{Related Work}
\label{sec:related-work}
The use of ML techniques in cosmology has been an area of great interest in the recent years, mainly due to the potential of these model to replace expensive computations.  \citet{Ramanah2019} implemented a convolutional GAN  capable of generating halo fields from cheap Lagrangian perturbation theory initial conditions. While this approach is significantly cheaper than an $N$-body simulation, is is entirely field-based,  thus it generates haloes  in a mesh rather than discrete haloes. The use of mesh-structured data in this case limits the spatial resolution of the output at the grid cell size but allows for a faithful reproduction of the large-scales. While the final product is somewhat similar to our method, we have the advantage of not being  limited by any  grid  spatial resolution.

The approach presented in \citet{Li2021, Ni2021} is more similar to what we aim at accomplishing with our method. \citet{Li2021} develops a convolutional GAN that generates large displacement fields that allow them to move initial particle data. While the displacements themselves are limited by the grid size, the spatial resolution of the particle data is not and the output of the model is a particle catalogue rather than a field. \citep{Ni2021} extends this model to produce not only displacements but also velocity fields. The output of their models is then used to create halo catalogues which agree at a $\sim10 {\rm per~cent}$ level with those of their HR reference. Moreover the authors quote $\sim10 {\rm per~cent}$ agreement with their reference in two and three point clustering measurements on the dark matter particles. The resolution difference between input and output is a factor 8 per dimension, going from $64^3$ to $512^3$ particles.

\section{Conclusions}
\label{sec:conclusions}
With the deployment of latest redshift surveys and the development of new and larger ones, the amount of cosmological data is expected to increase by orders of magnitude. The reduced statistical uncertainty that comes with larger data volumes may cause that systematical errors that can currently be safely ignored, become important in the future. This implies that large cosmological volumes must be simulated with a high mass resolution in order to be able to study these effects. However, the cost of full $N$-body simulations with the required size and resolution is quickly becoming too large for the latest hardware and software. To overcome this issue, we present a Machine Learning approach that generates an approximate high resolution halo catalogue by modifying the halo masses of a low resolution FoF halo catalogue, thereby avoiding the full computation of the high resolution simulation.

In this work we have presented a random forest regressor that predicts high resolution halo masses when given the dark matter overdensity, halo mass and halo particle velocity dispersion obtained from a low resolution halo catalogue. We base our method in matching pairs of low and high resolution haloes based on spatial distance and a mass threshold. We show that the halo mass function is reproduced within 3 {\rm per~cent} error for masses larger than $10^{12.5}~h^{-1}M_\odot$ and larger error for smaller masses (e.g., 10 {\rm per~cent} at $10^{11.5}~h^{-1}M_\odot$). This is an improvement compared to the original low-resolution catalogue of which HMF is biased by more than 3 {\rm per~cent} at $<10^{13.5}~h^{-1}M_\odot$. On the other hand, we argue that these errors in HMF is not important for real applications such as halo occupation distribution.
Moreover, our approach includes a classifier that is able to discard low resolution haloes that are not likely to have a high resolution halo match.

Even though our approach only predicts halo masses, the two-point clustering is recovered for haloes of masses larger than $\sim 2.5\times 10^{11} h^{-1}M_\odot$, which amounts to $\sim 200$ HR and 25 LR dark matter particles. These haloes are small enough to potentially host ELGs, which are a major tracer for galaxy clustering for DESI. The power spectrum of the prediction is shown to be within a 3 {\rm per~cent} error with respect to the real high resolution catalogue down to scales of $k\approx1~h\,\rm Mpc^{-1}$ in both real and redshift space. Moreover, the power spectrum is also recovered within 5 {\rm per~cent} error in various mass bins. In configuration space, the low-scales ($s\sim1 h^{-1}\rm Mpc$) show large discrepancies between the one-to-one matched and the real high resolution catalogues, which affect the prediction of our model. These discrepancies are therefore attributed to our matching algorithm rather than the model itself.

The predicted catalogue also shows three-point clustering consistent with the reference one within 10 {\rm per~cent}, which is a large improvement compared to the low resolution catalogue. The mass-binned bispectra are shown to be bias corrected as well, though there are still obvious variances with $\theta$ after the correction. 

We attribute the success of the model in recovering these various statistics to our unique one-to-one approach and the halo matching procedure we devised for creating the training data set. Indeed, there seems to be enough environmental information encoded by the matching such that the designed environmental halo features $\delta^{\rm M}_h$ were not fundamental for the random forests and are therefore discarded in our fiducial model. We also find that the halo discarding procedure does not need to be perfect given that interchanging a pair of low mass, low resolution haloes is unlikely to change the clustering very significantly.

We highlight that our approach is advantageous in terms of hardware requirements when compared with models based on deep learning techniques applied to mesh-based data. The random forest is able to take halo features agnostic to the halo's position, meaning that it is scalable and independent on mesh size or resolution. Nonetheless, a random forest of the characteristics we found to give good predictions requires a large amount of RAM ($\sim 60$Gb in this study). In addition, our model seems to be robust to random initialisation.

Further work is needed to devise a similar model that is able to use Rockstar \citep{Behroozi2013} halo features. Moreover, modern graph-based architectures are an interesting direction to be explored given the spatial structure of our data. Low resolution realizations may be created with a faster mesh-based N-body method such as FastPM \citep{Feng2016}, in which case the computational gain from a improved resolution method like ours will be even more significant.

\section*{Acknowledgements}
DFS acknowledges support from the SNF grant 200020\_175751. 
GY and SRT acknowledge financial support from the MICIU/FEDER (Spain) under project grant PGC2018-094975-C21.

The UNIT simulations used in this work  have been done in the MareNostrum Supercomputer at the Barcelona Supercomputing Center (Spain) thanks to the  cpu time awarded by PRACE under project grant number 2016163937.
FOF catalogues have been calculated at LRZ Munich within the project pr74no.

\section*{Data Availability}
The halo catalogues will be publicly available through the UNIT simulation website \url{ http://www.unitsims.org}.



\bibliographystyle{mnras}
\bibliography{refs} 




\appendix



\bsp	
\label{lastpage}
\end{document}